\documentclass{llncs}

\usepackage{soul}

\usepackage{comment} %
\renewcommand{\orcidID}[1]{\href{https://orcid.org/#1}{\includegraphics[scale=.03]{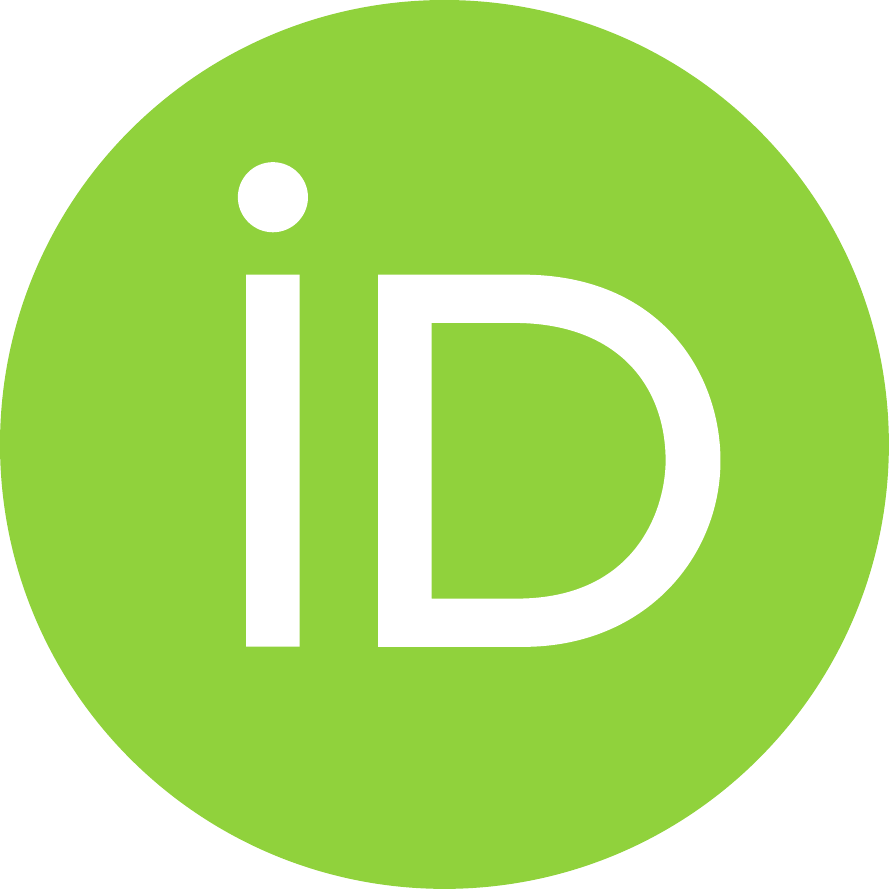}}}

\authorrunning{C. Alegr\'{i}a et al.}

\newif \ifVersioneShort
\VersioneShortfalse

\newif \ifCameraReady
\CameraReadytrue
\let\oldendproof\endproof
\renewcommand\endproof{~\hfill$\qed$\oldendproof}

\newcommand{\parametroApxproof}{
\ifVersioneShort
strip
\else
inline
\fi
}
\usepackage[bibliography=common,appendix=\parametroApxproof]{apxproof}

\usepackage{amsfonts,amsmath,amssymb,amsthm}
\usepackage{graphicx}
\usepackage{soul}
\usepackage{xcolor}
\usepackage{xspace}
\usepackage{paralist}
\usepackage{complexity}

\usepackage{subfigure}
\usepackage{hyperref}
\usepackage[capitalise]{cleveref}

\renewcommand{\emph}[1]{\textcolor{blue}{\em #1}\xspace}

\usepackage[color=yellow]{todonotes}
\usepackage{fullpage}

\def\qed{\ifmmode\squareforqed\else{\unskip\nobreak\hfil
\penalty50\hskip1em\null\nobreak\hfil\squareforqed
\parfillskip=0pt\finalhyphendemerits=0\endgraf}\fi}

\graphicspath{{fig/}{./fig/}}

\DeclareMathOperator{\skel}{sk}
\DeclareMathOperator{\expan}{exp}
\newcommand{\pert}[1]{\ensuremath{H_{#1}}}

\newcommand{\UIRFE}{{\sc UIRFE}\xspace}
\newcommand{\UIRFElong}{\textsc{Unit-length Inner-Rectangular Fixed-Embedding}\xspace}
\newcommand{\UIR}{{\sc UIR}\xspace}
\newcommand{\UIRlong}{{\sc Unit-length Inner-Rectangular}\xspace}
\newcommand{\UR}{{\sc UR}\xspace}
\newcommand{\URlong}{{\sc Unit-length Rectangular}\xspace}
\newcommand{\URFE}{{\sc URFE}\xspace}
\newcommand{\URFElong}{{\sc Unit-length Fixed Embedding}\xspace}
\newcommand{\algorithmGrep}{{\sc Rectan\-gu\-lar-holes Algorithm}\xspace}

\Crefname{observation}{Observation}{Observations}
\Crefname{algorithm}{Algorithm}{Algorithms}
\Crefname{section}{Sect.}{Sects.}
\Crefname{observation2}{Observation}{Observations}
\Crefname{lemma}{Lemma}{Lemmas}
\Crefname{lemma2}{Lemma}{Lemmas}
\Crefname{claim}{Claim}{Claims}
\Crefname{claimx}{Claim}{Claims}
\Crefname{figure}{Fig.}{Figs.}
\Crefname{figure}{Fig.}{Figs.}
\Crefname{enumi}{Condition}{Conditions}
\Crefname{property}{Property}{Properties}
\Crefname{property2}{Property}{Properties}
\Crefname{assumption}{Assumption}{Assumptions}
\Crefname{theorem}{Theorem}{Theorems}
\Crefname{theorem2}{Theorem}{Theorems}

\newtheorem{observation}{Observation}

\newtheorem{claimx}{Claim}

\newtheoremrep{theorem2}[theorem]{Theorem}
\newtheoremrep{lemma2}[lemma]{Lemma}
\newtheoremrep{observation2}[observation]{Observation}
\newtheoremrep{property2}[property]{Property}

\newcommand{\todounfinished}[2][]{\todo[color=red!70, noline, size=\small, #1]{Unfinished: {#2}}}

\newcommand{\todonote}[2][]{\todo[color=blue!20, inline, #1]{#2}}

\newcommand{\qedclaim}{\hfill $\blacksquare$}
\newenvironment{claimproof}{\noindent{\itshape Proof.}}{~\hfill \qedclaim \smallskip}

\begin{document}

\title{Unit-length Rectangular Drawings of Graphs\thanks{This research was partially supported by PRIN projects no.\ 2022ME9Z78 ``NextGRAAL'' and no.\  2022TS4Y3N ``EXPAND''.
A preliminary version of this research has appeared in the Proceedings of the 30th International Symposium on Graph Drawing and Network Visualization~\cite{AlegriaLBFGP22}.
}}

\author{
    Carlos Alegr\'{i}a\inst{}\orcidID{0000-0001-5512-5298} \and
    Giordano {Da Lozzo}\orcidID{0000-0003-2396-5174} \and
    Giuseppe {Di Battista}\orcidID{0000-0003-4224-1550} \and \\
    Fabrizio Frati\orcidID{0000-0001-5987-8713} \and
    Fabrizio Grosso\orcidID{0000-0002-5766-4567} \and
    Maurizio Patrignani\orcidID{0000-0001-9806-7411}}

\institute
{
  Department of Engineering, Roma Tre University, Rome, Italy\\
  \email{carlos.alegria@uniroma3.it},
  \email{giordano.dalozzo@uniroma3.it},
  \email{giuseppe.dibattista@uniroma3.it},
  \email{fabrizio.frati@uniroma3.it},
  \email{fabrizio.grosso@uniroma3.it},
  \email{maurizio.patrignani@uniroma3.it}
}

\maketitle

\begin{abstract}
A {\em rectangular drawing} of a planar graph $G$ is a planar drawing of $G$ in which vertices are mapped to grid points, edges are mapped to horizontal and vertical straight-line segments, and faces are drawn as rectangles. 
Sometimes this latter constraint is relaxed for the outer face.
In this paper, we study rectangular drawings in which the edges have unit length.
We show a complexity dichotomy for the problem of deciding the existence of a unit-length rectangular drawing, depending on whether the outer face must also be drawn as a rectangle or not.
Specifically, we prove that the problem is \NP-complete for biconnected graphs when the drawing of the outer face is not required to be a rectangle, even if the sought drawing must respect a given planar embedding, whereas it is polynomial-time solvable, both in the fixed and the variable embedding settings, if the outer face is required to be drawn as a rectangle.
Furthermore, we provide a linear-time algorithm for deciding whether a plane graph admits an embedding-preserving unit-length rectangular drawing if the drawing of the outer face is prescribed.
As a by-product of our research, we provide the first polynomial-time algorithm to test whether a planar graph~$G$ admits a~rectangular~drawing, for general instances of maximum~degree~$4$.

\end{abstract}

\section{Introduction}
Among the most celebrated aesthetic criteria in Graph Drawing we have: (i) planarity, (ii) orthogonality of the edges, (iii) unit length of the edges, and (iv) convexity of the faces. 
We focus on drawings in which all the above aesthetics are pursued at once.
Namely, we study orthogonal drawings where the edges have length one and the faces are rectangular.

Throughout the paper, any considered graph drawing has the vertices mapped at {\em distinct} points of the plane.
Orthogonal representations are a classic research topic in Graph Drawing. 
A rich body of literature is devoted to orthogonal drawings of planar~\cite{%
DBLP:journals/siamcomp/BattistaLV98,%
dlop-oodlt-20,%
DBLP:conf/gd/GargL99,%
DBLP:journals/siamdm/ZhouN08%
} and plane~\cite{%
DBLP:journals/jgaa/CornelsenK12,%
DBLP:journals/jgaa/RahmanNN99,%
DBLP:conf/wg/RahmanN02,%
DBLP:conf/stoc/Storer80,%
DBLP:journals/siamcomp/Tamassia87%
}
graphs with a minimum number of bends in total or per edge~\cite{%
DBLP:journals/comgeo/BiedlK98,%
DBLP:journals/algorithmica/Kant96,%
DBLP:journals/dam/LiuMS98%
}.
An orthogonal drawing with no bend is 
a \emph{rectilinear} drawing.
Several papers address rectilinear drawings of planar~\cite{%
DBLP:conf/compgeom/ChangY17,%
DBLP:conf/gd/Frati20,%
DBLP:journals/siamcomp/GargT01,%
DBLP:conf/cocoon/Hasan019,%
DBLP:conf/gd/RahmanEN05,%
DBLP:journals/ieicet/RahmanEN05%
} and plane~\cite{%
DBLP:conf/gd/Didimo0LO20,%
DBLP:conf/gd/Frati20,%
DBLP:journals/jgaa/RahmanNN03,%
DBLP:journals/siamcomp/VijayanW85%
} graphs.
When all the faces of a rectilinear drawing have a rectangular shape the drawing is \emph{rectangular}. Maximum degree-3 plane graphs admitting rectangular drawings were first characterized in~\cite{Th84,u-drm-53}.
A linear-time algorithm to find a rectangular drawing of a maximum degree-$3$ plane graph, provided it exists, is described in~\cite{DBLP:journals/comgeo/RahmanNN98} and extended to maximum degree-$3$ planar graphs in~\cite{DBLP:journals/jal/RahmanNG04}.
Surveys on rectangular drawings can be found in~\cite{%
f-rsrpg-13,%
DBLP:books/ws/NishizekiR04,%
nr-rda-13%
}.
If only the internal faces are constrained to be rectangular, then the drawing is called \emph{inner-rectangular}.
In~\cite{mhn-irdpg-06} it is shown that a plane graph $G$ has an inner-rectangular drawing $\Gamma$ if and only if a special bipartite graph constructed from $G$ has a perfect matching. Also, $\Gamma$ can be found in $O(n^{1.5}/\log n)$ time if $G$ has $n$ vertices and a ``sketch'' of the outer face is prescribed, i.e., all the convex and concave outer vertices are prescribed.

\begin{figure}[t!]
\includegraphics[width=\textwidth,page=4]{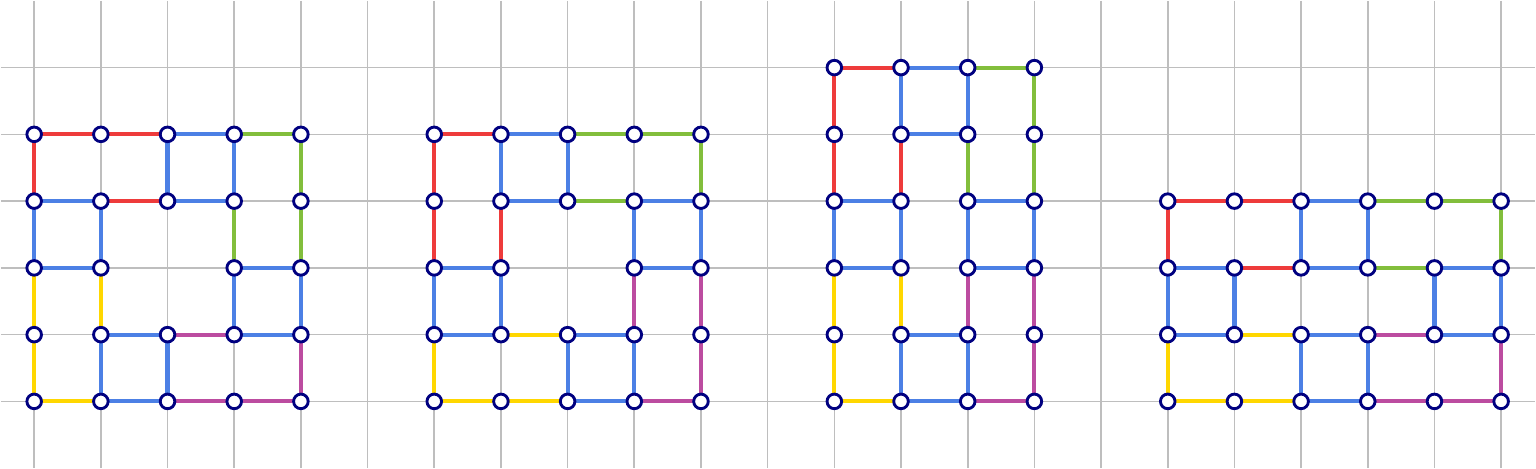}

\caption{
Unit-length embedding-preserving rectangular drawings of a plane graph.
}
\label{fig:grid-drawing-c}
\end{figure}

Computing straight-line drawings whose edges have constrained length is another core topic in graph drawing~\cite{%
DBLP:conf/compgeom/AbelDDELS16,%
DBLP:conf/gd/AlegriaBLBFP21,%
DBLP:conf/gd/AngeliniLBBHPR14,%
DBLP:conf/alenex/0001S20,%
cdr-pegse-07,%
DBLP:journals/dam/EadesW90,%
Schaefer2013%
}.
The graphs admitting planar straight-line drawings with all edges of the same length are also called \emph{matchstick graphs}.
Recognizing matchstick graphs is 
\NP-hard for biconnected~\cite{DBLP:journals/dam/EadesW90} and triconnected~\cite{cdr-pegse-07} graphs,
and in fact, even strongly
$\exists\mathbb{R}$-complete~\cite{DBLP:conf/compgeom/AbelDDELS16}; see also~\cite{Schaefer2013}.

A \emph{unit-length grid drawing} maps vertices to grid points and edges to horizontal or vertical segments of unit Euclidean length. A \emph{grid graph} is a graph that admits a unit-length grid drawing\footnote{Note that in some literature the term ``grid graph'' denotes an ``induced'' graph, i.e., there is an edge between any two vertices at distance one. See, for example, \cite{ips-hpgg-82}.}.
Recognizing grid graphs is \NP-complete for ternary trees of pathwidth~$3$~\cite{DBLP:journals/ipl/BhattC87}, for binary trees~\cite{g-ulebt-89}, and for trees of pathwidth~$2$~\cite{gsz-grcpc-21}, but solvable in polynomial time on graphs of pathwidth~$1$~\cite{gsz-grcpc-21}.
A variant of the grid graph recognition problem is when the drawing is constrained to be contained in the $k \times r$ grid.  In~\cite{gsz-grcpc-21} this problem is shown to be NP-hard when $k=3$ even on graphs of pathwidth~$2$. The same problem is shown to be fixed-parameter tractable (FPT) parameterized by $k + mcc$, where $mcc$ is the maximum size of a connected component of $G$.
An exponential-time algorithm to compute, for a given weighted planar graph, a rectilinear drawing in which the Euclidean length of each edge is equal to the edge weight has been presented in~\cite{DBLP:conf/alenex/0001S20}.

Let $G$ be a planar graph. The {\sc Unit-length Inner-Rectangular Drawing Recognition} (for short, \UIR) problem asks whether a unit-length inner-rectangular drawing of $G$ exists. Similarly, the {\sc Unit-length Rectangular Drawing Recognition} (for short, \UR) problem asks whether a unit-length rectangular drawing of $G$ exists.
Let now $H$ be a plane {\em or} planar embedded (i.e., no outer face specified) graph. The {\sc Unit-length Inner-Rectangular Drawing Recognition with Fixed Embedding} (for short, \UIRFE) problem asks whether a unit-length inner-rectangular embedding-preserving drawing of $H$ exists. Similarly, the {\sc Unit-length Rectangular Drawing Recognition with Fixed Embedding} (for short, \URFE) problem asks whether a unit-length rectangular embedding-preserving drawing of $H$ exists; \cref{fig:grid-drawing-c} shows different unit-length rectangular drawings of the same plane graph.

\paragraph{Our contribution.} 
In \cref{se:hardness} we show \NP-completeness for the \UIRFE (\cref{th:biconnected}) and \UIR (\cref{th:biconnected_uir}) problems when the input graph is biconnected, which is
surprising
since a biconnected graph has degrees of freedom that are more restricted than those of a tree.
In \cref{se:linear-time-algorithm-prescribed-drawing-external-face} we provide a linear-time algorithm for the \UIRFE and \URFE problems if the drawing of the outer face is given (\cref{th:grep}).
In \cref{se:unit-length-rectangular-drawings} we first show that the \URFE problem is cubic-time solvable; the time bound becomes linear if all internal faces of the input graph have maximum degree~$6$. These results hold both when the outer face is prescribed (\cref{th:rect-grep-outer-face}) and when it is not (\cref{th:rect-grep}).
Second, we show a necessary condition for an instance of the \UR problem to be positive in terms of its SPQR-tree (\cref{lem:structural}). Exploiting the above condition, we show that the \UR problem is cubic-time solvable; the running time becomes linear when the SPQR-tree of the input graph satisfies special conditions (\cref{th:linear-cubic-dichotomy}). Finally, as a by-product of our research, we 
provide the first polynomial-time algorithm to test whether a planar graph~$G$ admits a~rectangular~drawing, for general instances of maximum~degree~$4$ (\cref{th:non-unit-rectangular-drawing}).

\section{Preliminaries}
\label{se:preliminaries}

For basic graph drawing terminology and definitions refer, e.g., to~\cite{DBLP:books/ph/BattistaETT99,DBLP:books/ws/NishizekiR04}.

\begin{figure}[tb!]
\subfigure[\label{fig:grid-drawing-a}]{
\includegraphics[width=.5\textwidth,page=1]{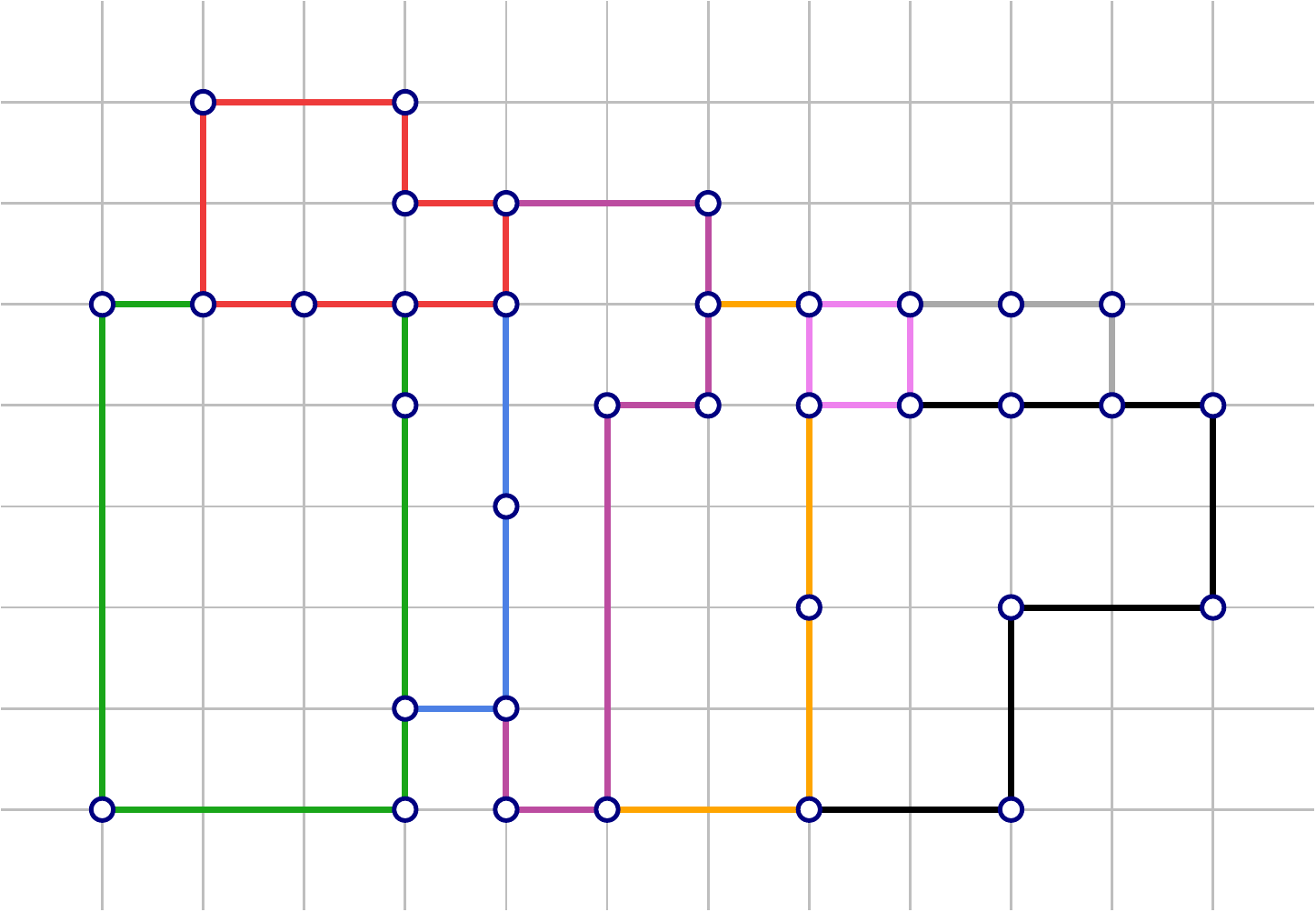}
}
\subfigure[\label{fig:grid-drawing-b}]{
\includegraphics[width=.5\textwidth,page=2]{fig/gr.pdf}
}
\caption{\subref{fig:grid-drawing-a} A planar rectilinear grid drawing of a graph. \subref{fig:grid-drawing-b} A unit-length rectangular grid drawing of the same graph.
}
\label{fig:grid-drawing}
\end{figure}

\paragraph{\bf Drawings and embeddings.}
A \emph{drawing} of a graph maps each vertex to a distinct point on the plane and each edge to a Jordan arc connecting its end-vertices.
A drawing of a graph is \emph{planar} if it contains no vertex-edge overlaps and no edge-edge crossings. A planar drawing of a graph partitions the plane into topologically connected regions called \emph{faces}.
The unbounded face is called the \emph{outer face}.
Two planar drawings of a connected graph are \emph{planar equivalent} if they induce the same counter-clockwise ordering of the edges incident to each vertex. Also, they are \emph{plane equivalent} if they are planar equivalent and the clockwise order of the edges along the boundaries of their outer faces is the same.
The equivalence classes of planar equivalent drawings are called \emph{planar embeddings}, whereas the equivalence classes of plane equivalent drawings are called \emph{plane embeddings}.
A \emph{planar embedded graph} is a planar graph equipped with one of its planar embeddings. Similarly,
a \emph{plane graph} is a planar graph equipped with one of its plane embeddings.
Given a planar embedded  (resp.~plane) graph $G$ and a planar (resp.~plane) embedding~$\mathcal E$ of~$G$, a planar drawing $\Gamma$ of $G$ is \emph{embedding-preserving}~if $\Gamma \in \mathcal{E}$. 
Consider a planar graph~$G$.
If $G$ is connected, a planar embedding of~$G$ is defined by the counter-clockwise circular order of the edges incident on each vertex, while a plane embedding of~$G$ is defined by the counter-clockwise circular order of the edges incident on each vertex and by the choice of the outer face. 
Instead, if $G$ is not connected, then a planar (plane) embedding is defined by a planar (plane) embedding of each of its connected components, together with the relative placement of each of these components to one another, called \emph{relative positions}, which specifies an assignment of each connected component to one face of each other~component.

\paragraph{\bf Geometric definitions}%
A \emph{polygon} is a closed polygonal chain consisting of a finite number of straight-line segments.
A polygon \emph{intersects itself} if two segments non-adjacent in the chain have a non-void intersection. 
A polygon is \emph{simple} if it does not intersect itself. This implies that there are no repeated segments or points in the chain.
A polygon is \emph{weakly simple} if it bounds a region of the plane that is homeomorphic to an open disk. A simple polygon is \emph{convex} if its interior is a convex set. A \emph{convex drawing} of a planar graph $G$ is a straight-line planar drawing of $G$ in which all the faces are drawn as convex polygons, including the outer face. In~\cite{DBLP:journals/iandc/BattistaTV01}, it has been shown that a planar graph admits a convex drawing only if it is biconnected.
A \emph{convex subdivision} of a simple polygon $P$ is partition of the interior of $P$ into convex sets. Note that a convex drawing defines a convex subdivision of the polygon bounding the outer face.
\paragraph{\bf Unit-length rectangular drawings.}
A \emph{rectilinear drawing} of a graph is a drawing such that each edge is an horizontal or vertical straight-line segment; see \cref{fig:grid-drawing-a}.
An \emph{inner-rectangular} drawing is a rectilinear drawing such that all its faces, except possibly for the outer face, are drawn as rectangles. An inner-rectangular drawing is \emph{rectangular} if its outer face is drawn as rectangle.
In a \emph{grid drawing}, vertices are mapped to points with integer coordinates (i.e., \emph{grid points}).
A drawing of a graph in which all edges have unit Euclidean length is a \emph{unit-length drawing} (see \cref{fig:grid-drawing-b} for an example).

\begin{observation} \label{obs:uniform-implies-rectilinear}\label{obs:grid-implies-planar}
A unit-length grid drawing is rectilinear and planar.
\end{observation}

\begin{observation} \label{obs:unit-length-rectangular}
A unit-length rectangular (or inner-rectangular) drawing is planar and it is a grid drawing, up to a rigid transformation.
\end{observation}

The following simple property has been proved in \cite[Lemma~1]{10.1007/s11227-016-1811-y}.
\begin{property} \label{pro:uniform-cycle-implies-pair-vertices}
Every cycle that admits a unit-length grid drawing has even length. %
\end{property}

Since (inner) rectangular drawings exist only for maximum-degree-$4$ graphs, in the remainder, we assume that all considered graphs satisfy this requirement. 

\paragraph{\bf Connectivity.}
Let $G$ be a graph. A \emph{cut-vertex} (resp. \emph{separation pair}) in a graph $G$ is a vertex (resp. a pair of vertices) whose removal disconnects $G$. Graph $G$ is \emph{biconnected} (\emph{triconnected}) if it has no cut-vertex (resp.\ no separation pair). A \emph{biconnected component} (or \emph{block}) of a graph~$G$ is a maximal (in terms of vertices and edges) biconnected subgraph of~$G$. A block is \emph{trivial} if it consists of a single edge and \emph{non-trivial} otherwise. 
A \emph{split pair} of $G$ is either a pair of adjacent vertices or a separation pair.
The \emph{components}~of~$G$ \emph{with respect to} a split pair $\{u,v\}$ are defined as follows.
If $(u,v)$ is an edge of $G$, then it is a component of $G$ with respect to $\{u,v\}$.
Also, let $G_1,\dots,G_k$ be the connected components of $G \setminus \{u,v\}$. The subgraphs of $G$ induced by $V(G_i) \cup \{u,v\}$, minus the edge $(u,v)$, are components of $G$ with respect to $\{u,v\}$, for $i=1,\dots,k$.

\subsection{SPQR-trees}\label{se:spqr-trees}

We provide details of the SPQR-tree data structure introduced by Di Battista and
Tamassia~\cite{DBLP:journals/algorithmica/BattistaT96,dt-opl-96} to handle all planar
embeddings of a biconnected planar graph $H$.  The SPQR-tree $T$ of $H$ represents a decomposition of $H$ into triconnected components along its split pairs.  
Each node $\mu$ of $T$ is associated with a graph, called \emph{skeleton of $\mu$}, and denoted by $\skel(\mu)$. The edges of $\skel(\mu)$ are either edges of $H$, which we call \emph{real edges}, or newly introduced edges, called \emph{virtual edges}.
The tree $T$ is initialized to a single node $\mu$, whose skeleton, composed only of real edges, is $H$. 
Consider a split pair $\{u, v\}$ of the skeleton of some node $\mu$ of $T$,
and let $H_1,\dots,H_k$ be the components of $H$ with respect to $\{u,v\}$ such that $H_1$ is not a single virtual edge and, if $k=2$, also $H_2$ is not a single virtual edge.  
We introduce a node $\nu$ adjacent to $\mu$ whose skeleton is the graph $H_1 + e_{\nu,\mu}$, where $e_{\nu,\mu} = (u,v)$ is a virtual edge; also, we replace the skeleton $\skel(\mu)$ of $\mu$ with the graph $\bigcup_{i \neq 1} H_i + e_{\mu,\nu}$, where $e_{\mu,\nu} = (u,v)$ is a virtual edge.  We say that $e_{\nu,\mu}$ is the \emph{twin virtual edge of} $e_{\mu,\nu}$, and vice versa. Applying this replacement iteratively produces a tree with more nodes but smaller skeletons associated with the nodes. Eventually, when no further replacement is possible, the skeletons of the nodes of $T$ are of four types: parallels of at least three virtual edges ($P$-nodes), parallels of exactly one virtual edge and one real edge ($Q$-nodes), cycles of exactly three virtual edges ($S$-nodes), and triconnected planar graphs ($R$-nodes). The \emph{merge} of two adjacent nodes $\mu$ and $\nu$ in $T$, replaces $\mu$ and $\nu$ in $T$ with a new node $\tau$ that is adjacent to all the neighbors of $\mu$ and $\nu$, and whose skeleton is $\skel(\mu) \cup \skel(\nu) \setminus \{ e_{\mu,\nu}, e_{\nu,\mu}\})$, where the end-vertices of $e_{\mu,\nu}$ and $e_{\nu,\mu}$ that correspond to the same vertices of $H$ are identified. By iteratively merging adjacent $S$-nodes, we eventually obtain the (unique) SPQR-tree data structure as introduced by Di Battista and Tamassia~\cite{DBLP:journals/algorithmica/BattistaT96,dt-opl-96}, where the skeleton of an $S$-node is a cycle. The crucial property of this decomposition is that a planar embedding of $H$ uniquely induces a planar embedding of the skeletons of its nodes and that, arbitrarily and independently, choosing planar embeddings for all the skeletons  uniquely determines an embedding of $H$. Observe that the skeletons of $S$- and $Q$-nodes have a unique planar embedding, that the skeleton of $R$-nodes have two planar embeddings (which are one the reflection of the other), and that $P$-nodes have as many planar embedding as the permutations of their virtual edges.
Consider a node $\nu$ and a virtual edge $e_{\nu,\mu}$ in $\skel(\nu)$. Among the subtrees of $T$ obtained by removing the arc $(\nu,\mu)$ from $T$, let $T_{\nu,\mu}$ be the one that contains $\nu$.
The \emph{expansion graph} $\expan{(e_{\nu,\mu})}$ of $e_{\nu,\mu}$ is the subgraph of $H$ obtained by iteratively merging all the nodes in $T_{\nu,\mu}$ and by removing the virtual~edge~$e_{\nu,\mu}$.

It is often convenient to orient the arcs of $T$ so that, in the resulting directed tree, one $Q$-node $\rho$ is a sink and all other nodes have exactly one outgoing arc. Such an orientation corresponds to rooting $T$ at $\rho$, and we call it a \emph{normal orientation} of $T$. The next definitions assume a normal orientation of $T$. For a node $\mu \neq \rho$ of $T$, the \emph{poles} of $\mu$ are the endpoints of the virtual edge $e_{\nu,\mu}$ of $\skel(\mu)$ where $\nu$ is the parent of $\mu$; whereas the poles of $\rho$ are the endpoints of its unique virtual edge.
Consider any plane embedding $\cal E$ of $H$ in which the real edge corresponding to $\rho$ is incident to the outer face. Then $\cal E$ yields a plane embedding $\mathcal{E}_\mu$ of the skeleton of each node $\mu$ of $T$ in which the poles of $\mu$ are also incident to the outer face of $\mathcal{E}_\mu$. This motivates the next  definitions.
Consider a node $\mu\neq \rho$. Also, let $u$ and $v$ be the poles of $\mu$. Let $\nu$ be the parent of $\mu$ and let $e_{\mu,\nu}$ be the virtual edge representing $\mu$ in $\skel(\nu)$. Let $\mathcal{E}_\mu$ be the restriction of $\mathcal{E}$ to $\expan{(e_{\mu,\nu})}$ and let $H_\mu$ be the corresponding plane graph.
Note that there exist exactly two faces of $\mathcal{E}$ that are incident to edges of the outer face of $H_\mu$. We call such faces the \emph{outer faces} of $\mathcal{E}_\mu$.
By convention, we call \emph{left outer face} $\ell(\mathcal{E}_{\mu})$ of $\mathcal{E}_{\mu}$ (\emph{right outer face} $r(\mathcal{E}_{\mu})$ of $\mathcal{E}_{\mu}$) the outer face that is delimited by the path obtained by walking in clockwise direction (resp.\ in counter-clockwise direction) from $u$ to $v$ along the boundary of the~outer~face~of~$\mathcal{E}_{\mu}$. 
The terms left outer face and right outer face come from the fact that we usually think about $\mathcal E_{\mu}$ as having the pole $u$ at the bottom and the other pole $v$ at the top. 

If $H$ has $n$ vertices, then $T$ has $O(n)$ nodes and the total number of virtual edges in the skeletons of the nodes of $T$ is in $O(n)$.  From a computational complexity perspective, $T$ can be constructed in $O(n)$ time~\cite{DBLP:conf/gd/GutwengerM00}.

\section{NP-completeness of the UIRFE and UIR problems}
\label{se:hardness}

In this section we show \NP-completeness for both the \UIRFE and \UIR problems when the input graph is biconnected. Observe that both problems clearly lie in \NP, as a certificate for a biconnected $n$-vertex input graph consists of an injective mapping from the vertices of the graph to the points of a grid whose sides have length bounded by $\frac{n}{2}$. In fact, it is possible to verify in polynomial-time whether such a mapping defines a unit-length rectangular drawing, and in the positive case whether it respects a given planar embedding. Therefore, in the remainder of the section, we will focus on establishing polynomial-time reductions to show the \NP-hardness of the problems.
We start with the following theorem.

\begin{theorem}\label{th:biconnected}
  The {\sc Unit-length Inner-Rectangular Drawing Recognition with Fixed Embedding} problem is \NP-com\-ple\-te, even for biconnected plane graphs whose internal faces have maximum size~$6$.
\end{theorem}

Let $\phi$ be a Boolean formula in conjunctive normal form with at most three literals in each clause.
We denote by $G_\phi$ the \emph{incidence graph} of $\phi$, i.e., the graph that has a vertex for each clause of $\phi$, a vertex for each variable of $\phi$, and an edge $(c,v)$ for each clause $c$ that contains the \emph{positive literal} $v$ or the \emph{negated literal}~$\overline{v}$.
The formula~$\phi$ is an instance of {\sc Planar Monotone 3-SAT} if $G_\phi$ is planar and each clause of $\phi$ is either positive or negative.
A \emph{positive clause} contains only positive literals, while a \emph{negative clause} contains only negated literals.
Hereafter, w.l.o.g., we assume that all the clauses of $\phi$ contain {\em exactly}~three~literals. 
In fact, a clause with less than three literals can be modified by duplicating one of the literals in the clause, without altering the satisfiability of $\phi$.
Note that this modification might turn $G_\phi$ into a planar multi-graph.

\begin{figure}[tb!]
  \centering
  \subfigure[$\Gamma_\phi$]{
  \label{fig:planar-3-sat-original} \includegraphics[width=0.45\textwidth,page=1]{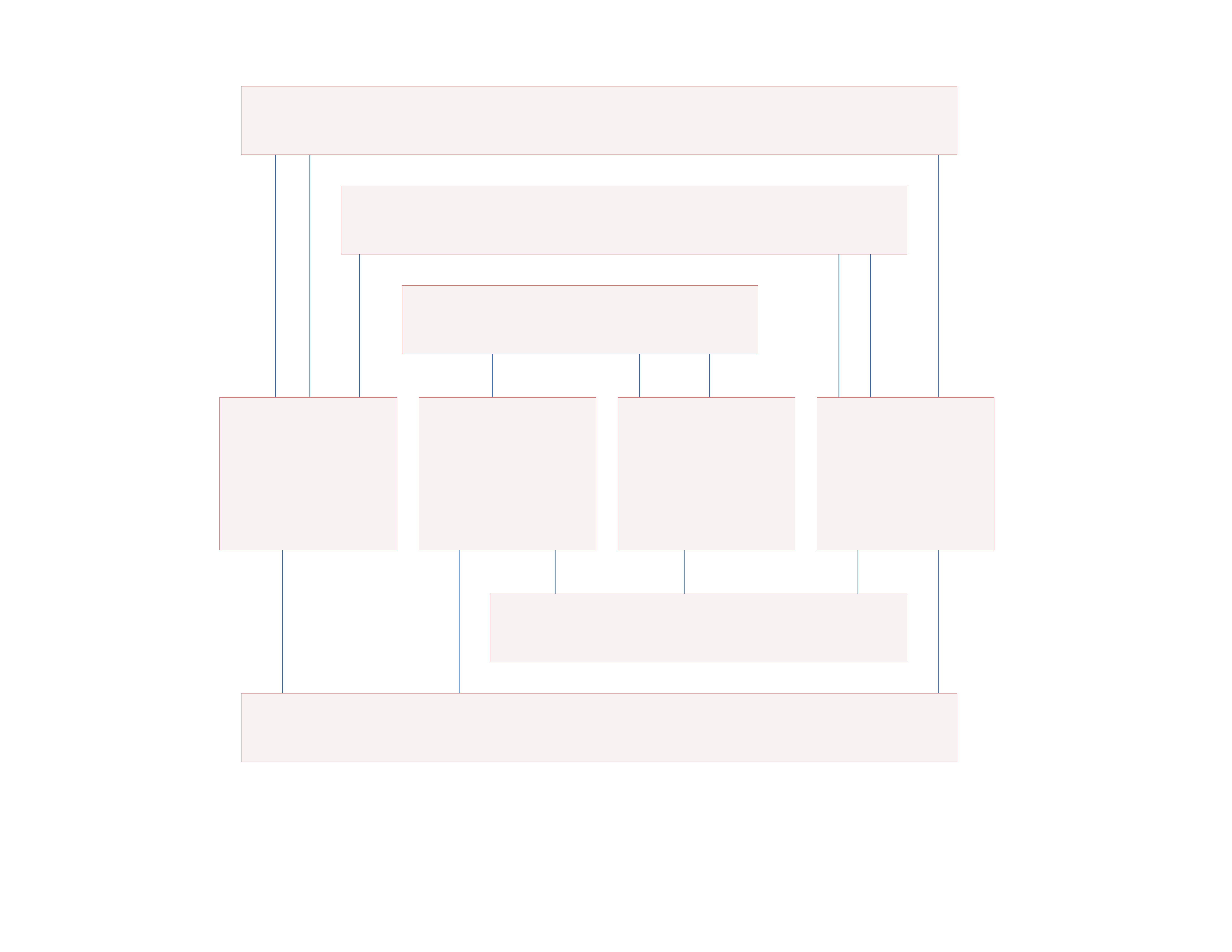}
  }
  \hfil
  \subfigure[$\Gamma^*_\phi$]{
  \label{fig:planar-3-sat-modified} \includegraphics[width=0.45\textwidth,page=2]{hardness_reduction_biconnected__overview}
  }
  \caption
  {
   \subref{fig:planar-3-sat-original} A  monotone rectilinear representation $\Gamma_\phi$ of $G_\phi$.
    The rectangles representing variables and clauses are red, whereas the line segments and rectangles representing the edges of $\phi$ are blue. 
\subref{fig:planar-3-sat-modified}
The auxiliary representation $\Gamma^*_\phi$ that satisfies the properties D\ref{prop:grid} to D\ref{prop:edges-representation}.}
\label{fig:rectilinear_representations}
\end{figure}

A \emph{monotone rectilinear representation} of $G_\phi$ is a drawing that satisfies the following properties (refer to \cref{fig:planar-3-sat-original}).
\begin{enumerate}[\bf {P}1:]
\item
  Variables and clauses are represented by axis-aligned rectangles with the same height.
\item
  The bottom sides of all rectangles representing variables lie on the same horizontal line.
\item
  The rectangles representing positive (resp.\ negative) clauses lie above (resp.\ below) the rectangles representing variables.
\item
  Edges connecting variables and clauses are represented by vertical segments.
\item
  The drawing is crossing-free.
\end{enumerate}

The {\sc Planar Monotone 3-SAT} problem is known to be \NP-complete, even when the incidence graph $G_\phi$ of $\phi$ is provided along with a monotone rectilinear representation $\Gamma_\phi$ of $G_\phi$~\cite{10.1142/S0218195912500045}.
We prove \cref{th:biconnected} by showing how to construct a plane graph $H_\phi$ that is biconnected, has internal faces of maximum size $6$, and admits a unit-length inner-rectangular drawing {\em if and only if} $\phi$ is satisfiable.
Our strategy is to modify $\Gamma_\phi$ to create a suitable auxiliary representation $\Gamma^*_\phi$  (see \cref{fig:rectilinear_representations}) and then to use the geometric information of $\Gamma^*_\phi$ as a blueprint to construct $H_\phi$.
We provide below a high-level description of the logic behind the reduction.

\subsection{The auxiliary monotone rectilinear representation $\Gamma^*_\phi$}
\label{ssse:hardness-biconnected:representation}

Hereafter, let $\delta^+_\phi$ (resp. $\delta^-_\phi$) be the maximum degree of $G_\phi$ when restricted to nodes representing variables and positive (resp. negative) clauses.
Let $\delta_\phi = \max(\delta^+_\phi, \delta^-_\phi)$. We denote by $|\phi|$ the size of $\phi$, that is, the number of variables plus the number of clauses in the formula. The auxiliary representation has the following properties (refer to
\cref{fig:rectilinear_representations}):

\begin{enumerate}[\bf {D}1:]
\item \label{prop:grid}
  
  The variables, clauses, and edges of $G_\phi$ are represented by axis-aligned rectangles whose corners have integer coordinates, i.e., they lie at grid points.
  
\item \label{prop:bounding-box}
  
  The width and height of the bounding box of $\Gamma^*_\phi$ are polynomially bounded in $|\phi|$.
  
\item \label{prop:variables-alignment}
  
  The rectangles representing variables have $O(\delta_\phi)$ width, constant height,
  and their bottom sides lie on a common horizontal grid~line.
  
\item \label{prop:clause-representation}

  Each rectangle representing a clause has $O(|\phi| \cdot \delta_\phi)$ width and constant height.

\item \label{prop:edges-representation}
  
  Each rectangle representing an edge has constant width and  $O(|\phi|)$ height.

\end{enumerate}

We can obtain $\Gamma^*_\phi$ by suitably translating and scaling the rectangles that represent the variables, clauses, and edges of $\phi$ in $\Gamma_\phi$.
Clearly, these transformations can be done in polynomial time in $|\phi|$.
We obtain the following lemma.

\begin{lemma}\label{lem:auxiliary-representation}
Starting from $\Gamma_\phi$, the representation $\Gamma^*_\phi$ can be constructed in polynomial time in $|\phi|$.
\end{lemma}

\begin{figure}[tb!]
  \centering
  \includegraphics[width=\textwidth,page=3]{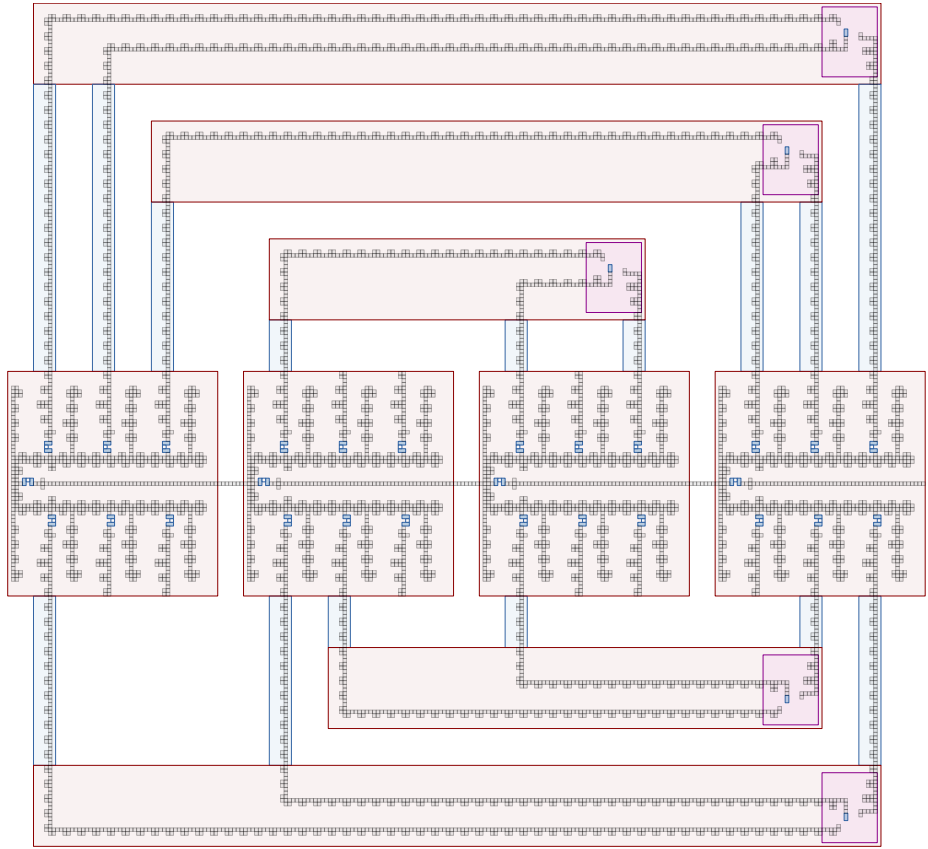}
  \caption{The graph $H_{\phi}$, with some faces omitted. Variable and clause gadgets are enclosed in light red boxes, while transmission gadgets are enclosed in light~blue~boxes.}
  \label{fig:biconnected_reduction_overview}
\end{figure}

\subsubsection{Overview of the reduction.}
\label{ssse:hardness-biconnected:overview}

The reduction is based on three main types of gadgets.
A variable $v \in \phi$ is modeled by means of a  \emph{variable gadget}, a clause $c \in \phi$ by means of an \emph{$(\alpha,\beta)$-clause gadget}, and an edge $(v,c) \in G_\phi$ by means of a \emph{$\lambda$-transmission gadget}.
We use the geometric properties of $\Gamma_\phi^*$ to determine the size and structure of each gadget, as well as how to combine the gadgets together to form $H_\phi$. In particular, we use the distances between the rectangles representing edges and clauses to compute the auxiliary parameters $\alpha$, $\beta$ and $\lambda$, which in turn are used to construct $(\alpha,\beta)$-clause gadgets and $\lambda$-transmission gadgets.
Finally, the incidences between the rectangles representing variables, edges, and clauses are used to decide how to join the edges of the gadgets to construct a biconnected graph.

\begin{figure}[p]
  \centering
  \subfigure[L-shape]{
  \label{fig:transmission_l-shape}
  {\includegraphics[scale=1,page=2]{hardness_reduction_biconnected__gadgets}}
  }
  \hfil
  \subfigure[Stick]{
  \label{fig:transmission_stick}
  {\includegraphics[scale=1,page=1]{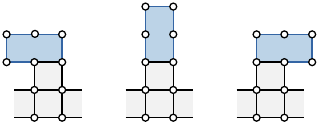}}
  }
  \hfil  
  \subfigure[C-shape]{
  \label{fig:transmission_c-shape}
  {\includegraphics[scale=1,page=3]{hardness_reduction_biconnected__gadgets}}
  }
  \caption{The unit-length grid drawings of the domino components. Frame faces are filled gray. Domino component faces are filled blue (size $6$) and white~(size~$4$).
  }\label{fig:domino_components}
\end{figure}

An example of a unit-length inner-rectangular drawing of $H_\phi$ is shown in \cref{fig:biconnected_reduction_overview}; some faces of $H_\phi$ are omitted. All these missing faces are part of \emph{domino components}, which admit a constant number of unit-length inner-rectangular drawings, see \cref{fig:domino_components}; some of these faces are shown filled in white or blue~in~\cref{fig:biconnected_reduction_overview}.

\begin{figure}[p]
  \centering
  \subfigure[A \texttt{true} configuration.]{\label{fig:biconnected_variable_gadget_a}\centering
  \includegraphics[width=0.49\textwidth,page=12]{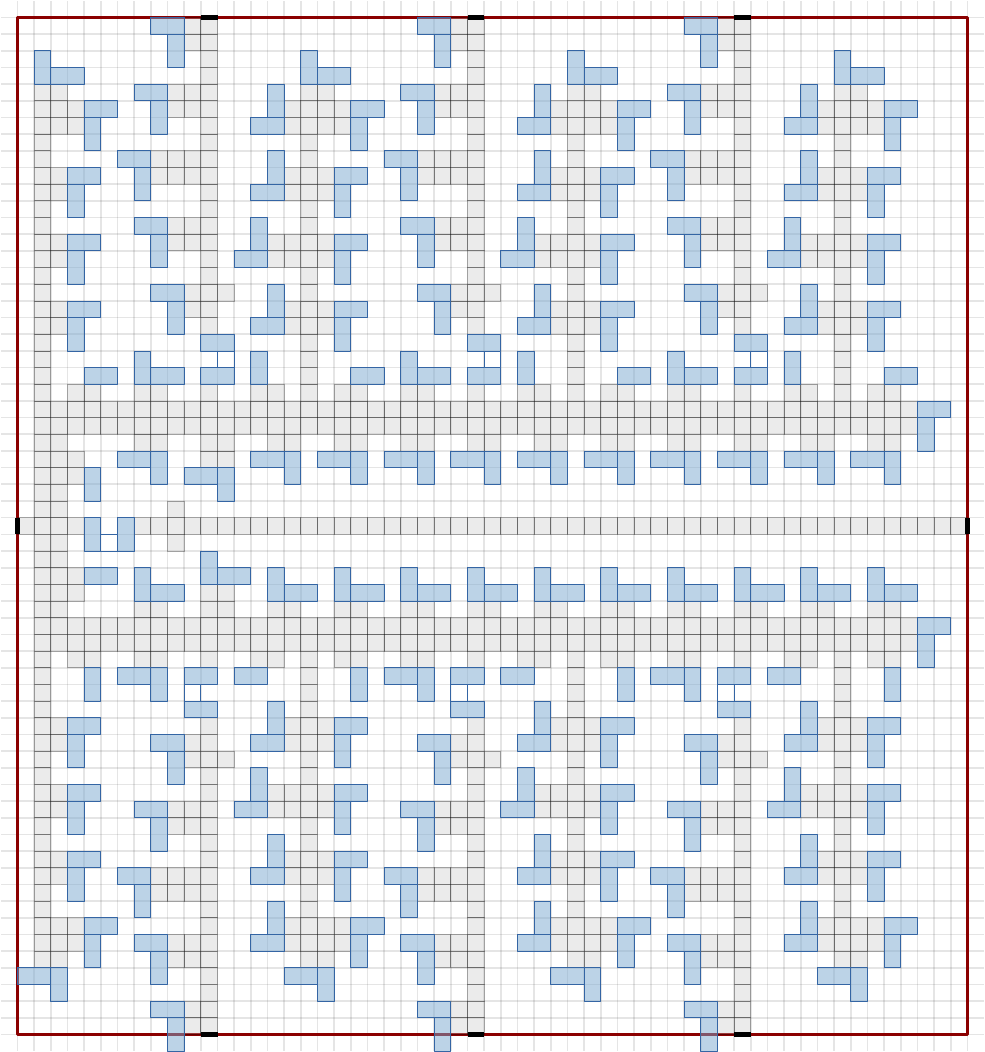}}
  \subfigure[A \texttt{false} configuration]{\label{fig:biconnected_variable_gadget_b}\centering
  \includegraphics[width=0.49\textwidth,page=13]{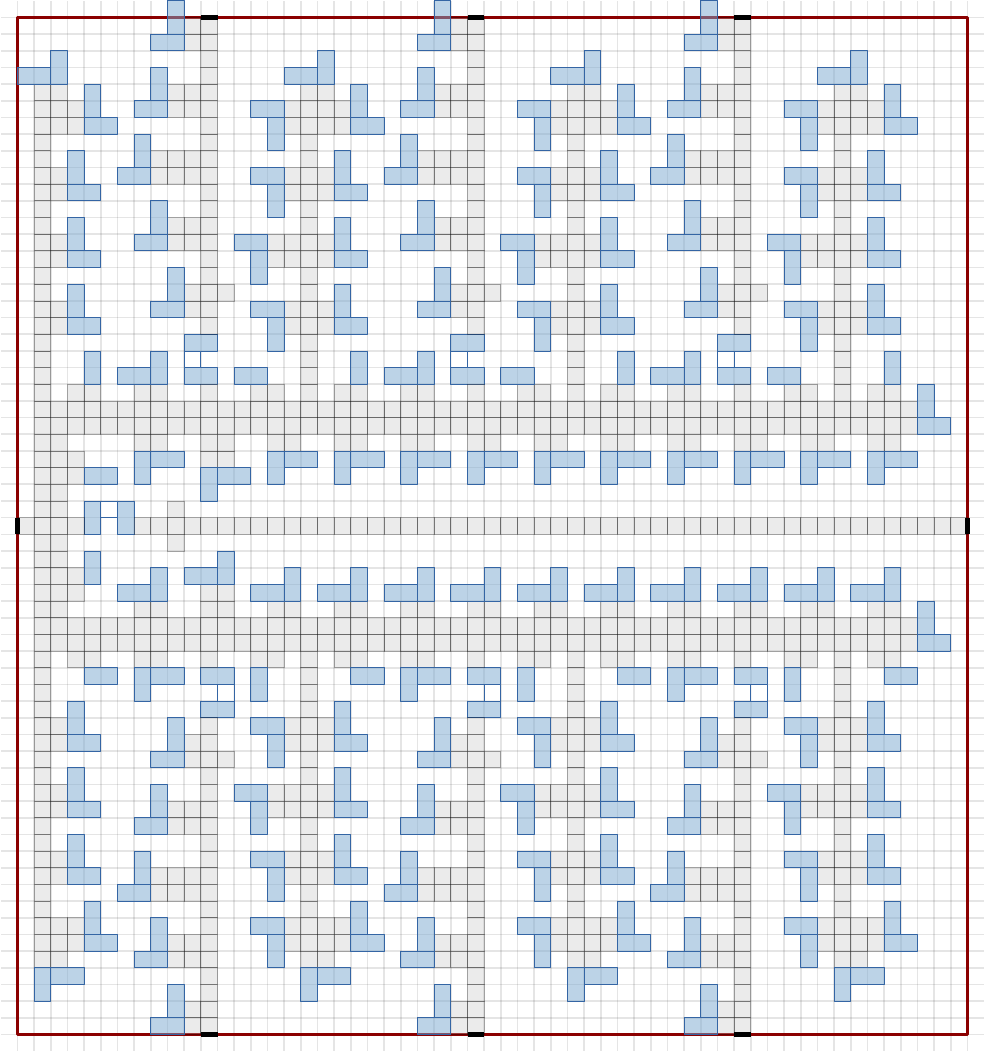}}
  \caption{The variable gadget.}
  \label{fig:biconnected_variable_gadget}
\end{figure}

Detailed illustrations of the variable, $(\alpha,\beta)$-clause, and $\lambda$-transmission gadgets are in \cref{fig:biconnected_variable_gadget,fig:biconnected_clause_gadget,fig:biconnected_transmission_gadget}, respectively.
In these figures, the enclosing rectangles of the gadgets are also included, as well as the faces of $H_\phi$ missing on \cref{fig:biconnected_reduction_overview}, which are shown in blue.

The logic behind the construction is as follows.
Consider the illustration of a variable gadget shown in \cref{fig:biconnected_variable_gadget}.
A variable gadget admits two unit-length inner-rectangular drawings.
These drawings differ from each other on whether the domino components cross the bottom (\cref{fig:biconnected_variable_gadget_a}) or the top (\cref{fig:biconnected_variable_gadget_b}) side of the red enclosing rectangle,
and correspond to a \texttt{true} and a \texttt{false} value assignment to the associated variable, respectively.
The truth assignments are propagated from variable to clause gadgets via $\lambda$-transmission gadgets.
The illustration of a $\lambda$-transmission gadget is shown in \cref{fig:biconnected_transmission_gadget}.
Consider the auxiliary interior purple rectangle~$R$.
A $\lambda$-transmission gadget is modeled in such a way that if a domino component sticking out of a variable gadget forces the bottom-most (resp.\ top-most) domino component to lie inside $R$, then there is a domino component crossing the top (resp.\ bottom) side of $R$.
In turn, the crossing domino component propagates the truth assignment by forcing a drawing of the domino components of the adjacent 
$(\alpha,\beta)$-clause gadget.
An illustration of an $(\alpha,\beta)$-clause gadget is shown in \cref{fig:biconnected_clause_gadget}.
An $(\alpha,\beta)$-clause gadget is designed in such a way that it admits a unit-length inner-rectangular drawing if and only if the drawings of its three incident $\lambda$-transmission gadgets allow for at least one domino component to cross the red rectangle.

\subsection{Description of the gadgets}
\label{ssse:hardness-biconnected:gadgets}

All the gadgets have internal faces of size either $4$ or $6$, and are formed by two sets of special subgraphs we call the \emph{frames} and the \emph{domino components}.
A frame is a biconnected subgraph formed by internal faces of size $4$, and has a unique unit-length inner-rectangular drawing (up to rigid transformations).
A domino component is instead a biconnected subgraph with internal faces of size either $4$ or $6$.
We define three different types of domino components: the L-shape, the C-shape, and the Stick.
Such components have constantly-many unit-length inner-rectangular drawings, shown in~\cref{fig:domino_components}. However, the geometry of the construction will force the L-shape components to have either the first or the last unit-length inner-rectangular drawing in~\cref{fig:transmission_l-shape} and the C-shape components to have either the first or the last unit-length inner-rectangular drawing in~\cref{fig:transmission_c-shape}.

\paragraph{Variable Gadget.}

Variable gadgets are formed by $2\delta_\phi + 2$ frames connected together by means of C-shape components, and a set of L-shape and stick components to propagate the truth assignment of the corresponding variable.
Refer to \cref{fig:biconnected_variable_gadget} for an illustration of the gadget.

Let $\mathcal{V}$ denote the variable gadget modeling some variable $v\in\phi$.
There are three crucial properties of the variable gadget.
First, C-shape components are adjacent to frames in such a way that in every unit-length inner-rectangular drawing of $\mathcal{V}$ the drawing of the frames of $\mathcal{V}$ is the same.
This implies that the bounding box $\mathcal{B}$ of the drawing of the frames of $\mathcal{V}$ does not change, regardless of the drawings of the C-shape components.  
Second, $\mathcal{V}$ admits two unit-length inner-rectangular drawings that we associate with the \texttt{true} (\cref{fig:biconnected_variable_gadget_a}) and \texttt{false} (\cref{fig:biconnected_variable_gadget_b}) truth assignments of $v$.
We remark that in the drawing corresponding to the \texttt{true} (resp. \texttt{false}) assignment, there are $\delta_\phi$ L-shape components crossing the bottom (resp. top) side of $\mathcal{B}$.
Finally, the gadget is constructed in such a way that the width and height of $\mathcal{B}$ are the same as those of the rectangle of $\Gamma_\phi^*$ representing $v$.

\paragraph{$\lambda$-transmission Gadget.}

\begin{figure}[tb!]
  \centering
  \subfigure[]
  {
    \label{fig:biconnected_transmission_gadget_a}
    \includegraphics[page=4]{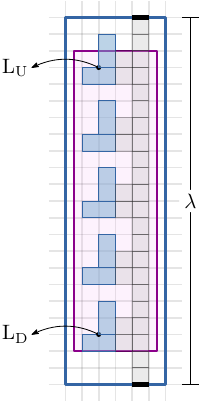}
  }
  \qquad
  \subfigure[]
  {
    \label{fig:biconnected_transmission_gadget_b}
    \includegraphics[page=5]{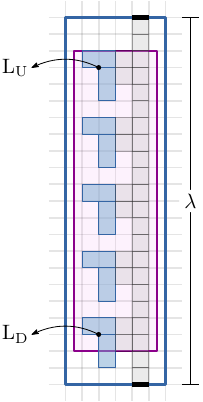}
  }
  \caption{Unit-length inner-rectangular drawings of a $\lambda$-transmission gadget for $\lambda=22$. \subref{fig:biconnected_transmission_gadget_a} If $L_D$ does not cross the purple rectangle, then $L_U$ crosses the purple rectangle. \subref{fig:biconnected_transmission_gadget_b} If $L_U$ does not cross the purple rectangle, then $L_D$ crosses the purple rectangle.}
  \label{fig:biconnected_transmission_gadget}
\end{figure}

The $\lambda$-transmission gadget is formed by a single frame, and a set of $\lfloor (\lambda - 2)/4 \rfloor$ L-shape components to propagate truth assignments from variable to clause gadgets.
Refer to \cref{fig:biconnected_transmission_gadget} for an illustration of the gadget.

Let $\mathcal{L}$ denote the $\lambda$-transmission gadget modeling some edge $(v,c)$ of $G_\phi$ that connects a variable $v$ to a clause $c$.
Consider the auxiliary purple rectangle $R$, and the L-shape components labeled with $L_D$ and $L_U$ in \cref{fig:biconnected_transmission_gadget}.
There are two crucial properties of the $\lambda$-transmission gadget.
First, in any unit-length inner-rectangular drawing of $\mathcal{L}$, if $L_D$ does not cross $R$ then $L_U$ crosses $R$, and vice versa.
Observe that, if an L-shape component of a variable gadget crosses the top (resp.\ bottom) side of its red enclosing rectangle, then $L_U$ (resp.\ $L_D$) crosses $R$.
This is how the truth assignment for a variable gets propagated through transmission gadgets.
Second, the width and height of the bounding box $\mathcal{B}$ of all the unit-length inner-rectangular drawings of $\mathcal{L}$ are the same.
Moreover, the width and height of $\mathcal{B}$ are less than or equal to the width and height of the rectangle of $\Gamma^*_\phi$ representing $(v,c)$.

\paragraph{$(\alpha,\beta)$-clause Gadget.}

\begin{figure}[tb!]
  \centering
  \subfigure[The gadget admits no unit-length inner-rectangular drawing in which no L-shape component crosses the red rectangle.]
  {
    \label{fig:biconnected_clause_gadget_a}
    \includegraphics[width=\textwidth,page=10]{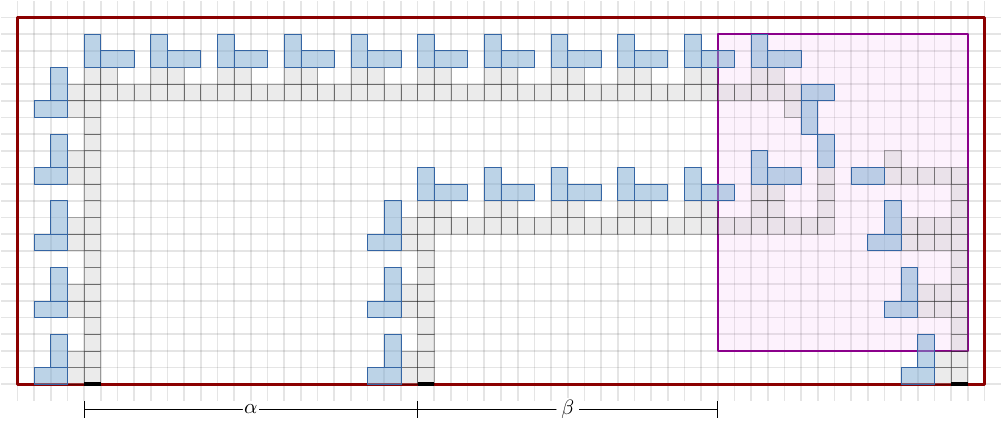}
  }
  \\
  \subfigure[The gadget admits a unit-length inner-rectangular drawing in which at least one L-shape component crosses the red rectangle.]
  {
    \centering
    \label{fig:biconnected_clause_gadget_b}
    \includegraphics[width=\textwidth,page=11]{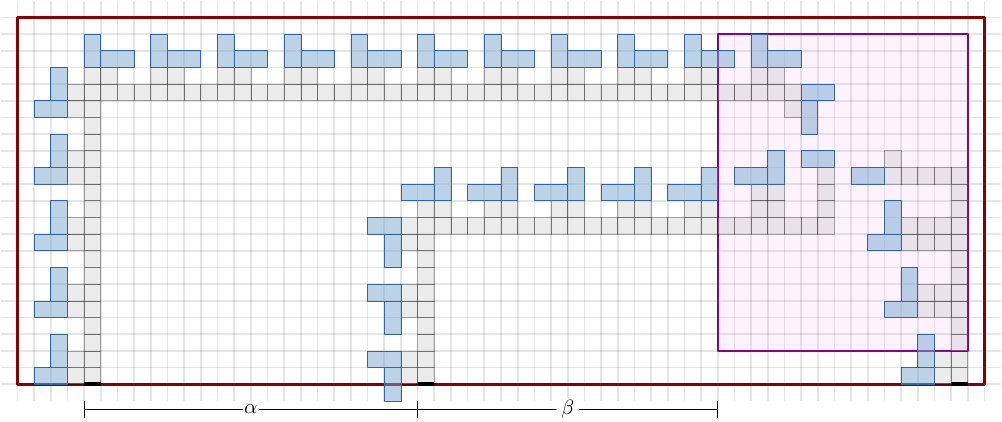}
  }

  \caption{The $(\alpha,\beta)$-clause gadget. The values of $\alpha$ and $\beta$ employed in the picture are smaller than they should be, for the sake of visibility.}
  \label{fig:biconnected_clause_gadget}
\end{figure}

In the following, refer to the example drawings of an $(\alpha,\beta)$-clause gadget shown in \cref{fig:biconnected_clause_gadget}.
Let $\mathcal{C}$ denote the $(\alpha,\beta)$-clause gadget modeling a clause $c\in\phi$.
Let $R$ denote the auxiliary purple rectangle shown in \cref{fig:biconnected_clause_gadget}.
The gadget $\mathcal{C}$ is formed by three disconnected components.
Each component is formed by a frame that, in the final graph $H_\phi$, is connected to the frame of a $\lambda$-transmission gadget modeling an edge of $G_\phi$ incident to $c$.
The components are also equipped with L-shape components to propagate the truth assignments coming from $\lambda$-transmission gadgets.

Consider for the moment the three connected subgraphs of $\mathcal{C}$ that admit a unit-length inner-rectangular drawing lying outside $R$.
Note that they are straightforward extensions of $\lambda$-transmission gadgets.
These auxiliary gadgets are used to propagate to $R$ the truth assignments coming from the boundary of the red enclosing rectangle.
Each auxiliary gadget has the property that, in any unit-length inner-rectangular drawing, if no L-shape component crosses the red enclosing rectangle, then there is one L-shape component crossing $R$.

Consider now the subgraphs of $\mathcal{C}$ that admit a unit-length inner-rectangular drawing lying in the interior of $R$.
The logic of the gadget is implemented by these subgraphs via the following crucial property: $\mathcal{C}$ admits a unit-length inner-rectangular drawing if and only if at least one L-shape component of the auxiliary gadgets is not crossing $R$.
See for example \cref{fig:biconnected_clause_gadget_a} in which all the three L-shape components of the auxiliary gadgets cross $R$, hence the $(\alpha,\beta)$-clause gadget does not admit a unit-length inner-rectangular drawing.

\subsection{Combining the gadgets together to form $H_\phi$}
\label{ssse:hardness-biconnected:combining}

For the purpose of combining two gadgets into a single connected graph, every gadget provides a set of special edges called \emph{attachment edges}.
In \cref{fig:biconnected_variable_gadget,fig:biconnected_clause_gadget,fig:biconnected_transmission_gadget}, the attachment edges are shown as thick black segments.
To combine two gadgets together, we first identify one attachment edge in each gadget, and then join the attachment edges together so that there is a single edge shared by both gadgets.

In the following description, the properties D\ref{prop:grid}-D\ref{prop:edges-representation} of $\Gamma^*_\phi$ are exploited to guarantee that, after combining all the gadgets, $H_\phi$ admits a unit-length inner-rectangular drawing if and only if $\phi$ is satisfiable.
To obtain $H_\phi$, we start by constructing the variable gadgets as above.
The variable gadgets are connected together by means of frames, each consisting of a sequence of a constant number of faces of size $4$, so that only faces that are consecutive in the sequence share vertices.
Each of such frames is combined with two variable gadgets by means of the attachment edges lying on the right and the left sides of their red enclosing rectangles.
The process continues by constructing a $\lambda$-transmission gadget for each edge of $G_\phi$.
The value of the parameter $\lambda$ of each gadget is the height of the blue rectangle of $\Gamma_\phi^*$ representing the associated edge of $G_\phi$.
A $\lambda$-transmission gadget and a variable gadget are combined together joining an attachment edge lying on the top (resp.\ bottom) side of the red enclosing rectangle of the variable gadget, and the attachment edge lying on the bottom (resp.\ top) side of the blue enclosing rectangle of the $\lambda$-transmission gadget.
We finally construct an $(\alpha,\beta)$-clause gadget for each clause of $\phi$.
We select the parameters $\alpha$ and $\beta$ according to the width of the red rectangles representing clauses in $\Gamma_\phi^*$, and the horizontal distances between the blue rectangles representing the edges incident to the modeled clause.
A $\lambda$-transmission gadget and an $(\alpha,\beta)$-clause gadget are combined together joining an attachment edge lying on the bottom (resp.\ top) side of the red enclosing rectangle of the $(\alpha,\beta)$-clause gadget, and the attachment edge lying on the top (resp.\ bottom) side of the blue enclosing rectangle of the $\lambda$-transmission gadget.

By the construction described above, it is not hard to see that $H_\phi$ is biconnected and admits a unit-length inner-rectangular drawing that preserves the given plane embedding if and only if $\phi$ is satisfiable.
The crucial property is that the domino components we use are forced to admit a constant number of unit-length inner-rectangular drawings that are all embedding preserving; see~again~\cref{fig:domino_components}.

By showing that the graph $H_\phi$ only admits unit-length inner-rectangular drawings that preserve the same plane embedding, we get the following.

\begin{theorem}\label{th:biconnected_uir}
  The \UIRlong problem is \NP-com\-ple\-te, even for biconnected planar graphs that admit an embedding in which the internal faces have maximum size $6$.
\end{theorem}

\begin{proof}
We show that the graph $H_\phi$ only admits unit-length inner-rectangular drawings that preserve the same plane embedding. 
Let $\Gamma$ and $\Gamma'$ be two unit-length inner-rectangular drawings of $H_\phi$.
By construction, every edge of $H_\phi$ belongs to at least a length-$4$ or a length-$6$ chordless cycle. Each such a cycle must necessarily bound an internal face of $\Gamma$ and $\Gamma'$. Therefore, the cycle bounding the outer face of $\Gamma$ and $\Gamma'$ is the same.
Let us call the cycles bounding the internal faces of $\Gamma$ and $\Gamma'$ the \emph{inner cycles} of $H_\phi$. 
By construction, any inner cycle shares at least an edge with another inner cycle. Therefore, the orientation of the inner cycles is the same in $\Gamma$ as in $\Gamma'$ (up to a reflection of the entire drawing). 
Therefore, all the unit-length inner-rectangular drawings admitted by $H_\phi$ preserve the same plane embedding, which is unique up to reflections of the whole drawing.
\end{proof}

Since any unit-length grid drawing of a cycle with $4$ or $6$ vertices is a rectangle, the previous theorem
implies the following result.

\begin{corollary} \label{cor:hardness-grid}
It is \NP-complete to decide whether a biconnected planar graph $G$ admits a unit-length grid drawing, even if $G$ has a prescribed plane embedding.
\end{corollary}

\begin{comment}
\subsection{From biconnected graphs to Trees and Caterpillars}\todounfinished{}
\label{sse:hardness-trees}

\begin{theorem}\label{th:trees}
  The \UIRFE problem is \NP-com\-ple\-te for plane trees.
\end{theorem}

\begin{proof}\todounfinished{}
  \todonote{Carlos: Given a biconnected graph $G$ with internal faces of maximum size $6$, construct a tree $T$ such that, $T$ has a unit-length grid drawing if and only if $G$ has a unit-length grid drawing. \cref{th:trees} then follows from \cref{th:biconnected}. Could we use somehow the SPQR Tree to substitute nodes of every type, one at a time?}
\end{proof}

%

%
%
%
%
%

%
%
%
%
%
%
%

%
%
%
%
%
%
%
%
%
%
%

%
%
%
%
%
%
%
%
%
%
%
%
%
%

%
%
%
%
%
%
%
%
%
%
%
%
%
%

%
%
%
%
%
%
%
%
%
%
%
%
%
%
%
%
%

\begin{theorem}
 \label{thm:binary_trees}
 The \UIRFE problem is \NP-complete for plane binary trees.
\end{theorem}
\begin{proof}\todounfinished{}
  The gadgets used for the reduction in \ref{th:trees} can be modified using a gadget from \cite{10.1016/0020-0190(89)90118-X} so that the tree is binary.
\end{proof}

\begin{theorem}
 \label{thm:caterpillar}
 The \UIRFE problem is \NP-complete for plane caterpillars.
\end{theorem}
\begin{proof}
  \todounfinished{}
  The proof is by reduction from the PLANAR MONOTONE 3-SAT problem.
  Extending the reduction we used in Theorem~\ref{thm:binary_trees}. 
\end{proof}

%
%
%

%
%
%
%
%
%

%
%
%
%
%
%
%
%
%
%
%

%
%
%
%
%
%

%
%
%

%
%
%
%
%
%
%
%
%

%
%
%
%
%
%
\end{comment}

%
%
%
%
%
%

%

%

%
%
%
%
%
%
%

%
%
%
%

%
%
%
%
%
%
%

%
%
%

%

%

%
%
%
%

%
%
%
%
%
%
%

%
%

%

%
%
%
%
%
%

%
%
%
%

%
%
%
%
%
%

%
%

\section{An Algorithm for the UIRFE and URFE Problems with a Prescribed Drawing of the Outer Face}\label{se:linear-time-algorithm-prescribed-drawing-external-face}
In this section, we show a linear-time algorithm for the \UIRFE (and consequently for the \URFE) problem in the case in which the drawing of the outer face is prescribed.

We start with two auxiliary lemmata.
The first one is an extension of a classical result by Devillers et al.~\cite{10.1016/S0925-7721(98)00039-X}.

\begin{lemma}\label{lem:planarity}
  Let $G$ be a connected planar graph and $\cal E$ be a plane embedding of $G$. A straight-line drawing $\Gamma$ of $G$ is planar and respects $\cal E$ if and only if:
  \begin{itemize}    
  \item
    for every face $f$ of $\mathcal E$, the walk delimiting $f$ is represented in $\Gamma$ by a weakly simple polygon, whose orientation is as prescribed by $\mathcal E$;
  \item
    for every vertex $v$ of $G$, the clockwise order of the edges incident to $v$ in $\Gamma$ is the same as in  $\mathcal E$; and
  \item
    let $C_o$ be the walk delimiting the outer face $f_o$ of~$\mathcal E$, and let $\Gamma_o$ be the weakly simple polygon representing $C_o$ in $\Gamma$; then every edge not in $C_o$ that is incident to a vertex $v$ of $C_o$, leaves $v$ towards the interior of~$\Gamma_o$.
  \end{itemize}
\end{lemma} 

\begin{proof}
The necessity is obvious. We prove the sufficiency. Suppose hence that $\Gamma$ satisfies the properties described in the statement. We prove that it is planar and respects $\mathcal E$. 

Let $\rho_h$ be equal to $3n-h-3$. Note that $\rho_h$ is the number of edges of a biconnected internally-triangulated $n$-vertex plane graph whose outer face is delimited by a cycle with $h$ vertices. We prove the lemma by induction on $\rho_k - |E(G)|$, where the index $k$ is the number of vertices of the convex hull of $\Gamma_o$.

If $\rho_k - |E(G)| = 0$, then each internal face of $\cal E$ is delimited by a $3$-cycle and $C_o$ is a $k$-cycle.
Since (i) each internal face of $\cal E$ is 
delimited in $\Gamma$ by a triangle, and $C_o$ is represented in $\Gamma$ by a convex $k$-gon, and since (ii) the clockwise order of the edges incident to each vertex in $\Gamma$ is as prescribed by $\cal E$, a classic result by Devillers et al.~\cite[Lemma~19]{10.1016/S0925-7721(98)00039-X} implies that $\Gamma$ is planar and induces a convex subdivision $\Gamma$ of $\Gamma_o$ (that also respects the planar embedding of $G$ obtained by disregarding the choice of the outer face of $\cal E$). Finally, the fact that $G$ is connected and that every edge not in $C_o$ that is incident to a vertex of $C_o$ leaves this vertex toward the interior of $\Gamma_o$ implies that $\Gamma_o$ bounds the outer face of $\Gamma$, and thus $\Gamma$ respects (the plane embedding) $\cal E$.

Let us now consider the case in which $\rho_k-|E(G)| > 0$. 
Let $f$ be a face of $\cal E$ such that either $f$ is an internal face of $\cal E$ of length at least $4$ or $f = f_o$ if the polygon $\Gamma_o$ bounding $f_o$ in $\Gamma$ is not convex. Then it is possible to draw in $f$ a straight-line segment $\overline{uv}$ between some pair of vertices $u$ and $v$ incident to $f$, such that $\overline{uv}$ does not cross any edge of $f$. In particular, $\overline{uv}$ divides $f$ into two faces $f'$ and $f''$. Let $G'$ be the plane graph obtained from $G$ by introducing the edge $(u,v)$ so that it splits the face $f$ into the faces $f'$ and $f''$. To define a plane embedding $\cal E'$ for $G'$,  it only remains to specify a choice for its outer face.  If $f \neq f_o$, then $f_o$ is the outer face of $\cal E'$. Otherwise, the outer face of $\cal E'$ is the unbounded face between $f'$ and $f''$. Let $\Gamma'$ be the drawing of $G'$ obtained from $\Gamma$ by drawing the edge $(u,v)$ as the straight-line segment $\overline{uv}$.

Observe that $\rho_k - |E(G')| < \rho_k - |E(G)|$. Furthermore, all the conditions of the statement are satisfied by $\Gamma'$, $\cal E'$, and $G'$. Therefore, by induction, $\Gamma'$ is planar and respects $\cal E'$. The fact that the restriction of $\Gamma'$ to $G$ yields a planar drawing $\Gamma$ of $G$ that respects $\cal E$ concludes the proof.%
\end{proof}

\begin{lemma}\label{lem:unicity}
Let $G$ be a plane graph and let $\Gamma_o$ be a unit-length grid drawing of the outer face $f_o$ of $G$. Then, an embedding-preserving inner-rectangular unit-length drawing of $G$ in which $f_o$ is delimited by $\Gamma_o$, if any, is unique.
\end{lemma}

\begin{proof}
First, we can assume that $G$ is connected. Indeed, if it is not, then the relative positions of distinct connected components of $G$ are such that each connected component lies in the outer face of each other, as otherwise obviously $G$ would not admit any inner-rectangular drawing and there would be nothing to prove. It follows that $\Gamma_o$ specifies a unit-length grid drawing of the outer face of each connected component of $G$ and thus it suffices to prove the lemma for an individual connected component of $G$ in order to prove it for $G$ itself. So in the rest of the proof we assume that $G$ is connected. We denote by $b(f)$ the walk of $G$ that bounds a face $f$. Note that, for any internal face $f$, $b(f)$ must be a simple cycle, as otherwise $G$ does not have a unit-length embedding-preserving inner-rectangular drawing.

We prove the lemma by induction on the number $i$ of internal faces of $G$. 
If $i=1$, then $G$ coincides with the cycle $b(f_o)$ and it admits a unit-length embedding-preserving inner-rectangular drawing if and only if $\Gamma_o$ is a rectangle oriented as prescribed by the embedding of $G$.

If $i>1$, then consider a vertex $v$ incident to $f_o$ with minimum $x$-coordinate in $\Gamma_o$. Let~$f$ be any internal face of $G$ incident to $v$ and let $P_{\textrm{left}}$ be the subgraph of $G$ induced by the vertices of $G$ with minimum $x$-coordinate in $\Gamma_o$. Observe that, if $P_{\textrm{left}}$ is not a collection of (chordless) paths, then $G$ does not admit a
unit-length embedding-preserving inner-rectangular  drawing
in which $f_o$ is delimited by $\Gamma_o$, and the statement trivially holds.
Let now $P_{\textrm{left}}(f) := P_{\textrm{left}} \cap b(f)$ be the subgraph of $P_{\textrm{left}}$ induced by the vertices on the boundary of $f$. 
If $P_{\textrm{left}}(f)$ consists of multiple connected components, then $f$ cannot be drawn as a rectangle in any unit-length embedding-preserving inner-rectangular  drawing of $G$ in which $f_o$ is delimited by $\Gamma_o$, and the statement trivially holds. In fact, since $v \in P_{\textrm{left}}(f)$, we have that the drawing of the left side of a rectangle $R$ representing $f$ must coincide with the drawing of $P_{\textrm{left}}(f)$. This in turn implies that $R$ is {\em prescribed}, that is, the rectangle $R$ representing $b(f)$ in an embedding-preserving inner-rectangular unit-length drawing of $G$ in which $f_o$ is delimited by $\Gamma_o$ is univocally determined.  Clearly, $G$ does not admit a unit-length embedding-preserving inner-rectangular drawing in which $f_o$ is delimited by~$\Gamma_o$ if {\bf (F1)} $R$ places a vertex in $V(f) \setminus V(f_o)$ on top of vertices in $V(f_o) \setminus V(f)$ or if {\bf (F2)} $R$ assigns a vertex on $V(f) \cap V(f_o)$ different coordinates than the ones prescribed by $\Gamma_o$.
If any of such conditions holds, then the statement trivially holds. Suppose now that neither (F1) nor (F2) occurs, and let $\Gamma'_o$ be the drawing obtained from $\Gamma_o$ by removing the edges of $P_{\textrm{left}}(f)$ and all the resulting isolated vertices, if any. Similarly, let $G'$ be the plane graph obtained by removing from~$G$ all the edges of $P_{\textrm{left}}(f)$ and all the resulting isolated vertices, if any.  Note that $G'$ is the plane subgraph of $G'$ whose internal faces are the faces of $G$ different from $f$ and whose outer face~$f'_o$ is obtained by merging $f_o$ and $f$, which is achieved by removing the edges and vertices of $P_{\textrm{left}}(f)$ except for its end-vertices. Also, note that $\Gamma'_o$ is a unit-length grid drawing of $f'_o$. Therefore, since $G'$ contains $i-1$ internal faces, we can now apply induction. The following two cases are possible. {\bf Case~1:} $G'$ does not admit 
a unit-length embedding-preserving inner-rectangular  drawing in which $f'_o$ is delimited by $\Gamma'_o$. In this case, $G$ does not admit 
a unit-length embedding-preserving inner-rectangular drawing in which $f_o$ is delimited by $\Gamma_o$, and the statement holds. {\bf Case~2:} Let $\Gamma'$ be the unique unit-length embedding-preserving inner-rectangular drawing
of $G'$ in which $f'_o$ is delimited by $\Gamma'_o$; note that, since we are not in {\bf Case 1}, such a drawing exists and is unique by the inductive hypothesis. Clearly, by adding $R$ to $\Gamma'$ we obtain
a unit-length embedding-preserving inner-rectangular drawing
of $G$ in which $f_o$ is delimited by $\Gamma_o$, which is unique since the drawing of $R$ is prescribed and since $\Gamma'$ is unique. This concludes the proof.%
\end{proof}

Consider a connected instance of the \UIRFE problem, i.e., an $n$-vertex  connected plane graph $G$; let $\cal E$ be the plane embedding prescribed for $G$. Let $\Gamma_o$ be a unit-length grid drawing of the walk bounding the outer face $f_o$ of $\cal E$.
W.l.o.g, assume that the smallest $x$- and $y$- coordinates of the vertices of $\Gamma_o$ are equal to~$0$.
Next, we describe an $O(n)$-time algorithm, called \algorithmGrep, to decide whether $G$ admits a unit-length inner-rectangular drawing that respects~$\cal E$ and in which the walk bounding $f_o$ is represented by~$\Gamma_o$.

We first check whether each internal face of $\cal E$ is bounded by a simple cycle of even length, as otherwise the instance is negative by \cref{pro:uniform-cycle-implies-pair-vertices}.
This can be trivially done in $O(n)$ time.
We remove from $G$ the bridges incident to the outer face and the resulting isolated vertices.

Now the algorithm processes the internal faces of $G$ one at a time.
When a face~$f$ is considered, the algorithm either detects that $G$ is a negative instance or assigns $x$- and $y$- coordinates to all the vertices of $f$.
In the latter case, we say that $f$ is \emph{processed} and its vertices are \emph{placed}.
Since the drawing of $f_o$ is prescribed, at the beginning~each vertex incident to $f_o$ is placed, while the remaining vertices are not. Also, every internal face of $\mathcal E$ is not processed.
The algorithm concludes that the instance is negative if one of the following conditions holds:
\begin{inparaenum}
[\bf (C1)]
\item \label{cond:vertex-different-coordinates}there is a placed vertex to which the algorithm tries to assign coordinates different from those already assigned to it, or
\item \label{cond:vertex-same-coordinates}there are two placed vertices with the same $x$-coordinate and the same $y$-coordinate.
\end{inparaenum}
If neither Condition C\ref{cond:vertex-different-coordinates} nor C\ref{cond:vertex-same-coordinates} occurs, after processing all the internal faces the vertex placement provides a unit-length inner-rectangular drawing of the input instance.

To process faces, the algorithm maintains some auxiliary data structures:
\begin{itemize}
\item
    A {\bf graph $H$}, called the \emph{\sc current graph}, which is the subgraph of $G$ composed of the vertices and of the edges incident to non-processed (internal) faces.
    Initially, we have $H=G$.
    In particular, we will maintain the invariant that each biconnected component of $H$ is non-trivial.
    We will also maintain the outer face of the restriction $\mathcal E_H$ of $\mathcal E$ to $H$, which we will still denote~by~$f_o$. When the current graph is $H$, all the vertices incident to the outer face of $\mathcal E_H$ are already placed, i.e., the drawing of each cycle delimiting the outer face of a biconnected component of $H$ is determined.
\item
    An {\bf array $A$}, called the \emph{\sc current outer-sorter}, that contains $M_x+1$ buckets, each implemented as a double-linked list, where $M_x$ is the largest $x$-coordinate of a vertex in $\Gamma_o$. %
    The bucket $A[i]$ contains the placed vertices of $H$ (i.e., those incident to the outer face of $H$) whose $x$-coordinate is~equal~to~$i$.
    Moreover, $A$ is equipped with the index $x_{\textrm{min}}$ of the first non-empty bucket.
    To allow removals of vertices in $O(1)$ time, we enrich each placed vertex with $x$-coordinate $i$ with a pointer to the corresponding list-item in the~list~$A[i]$.
    \item A {\bf set of pointers} for the edges of~$H$: Each edge $(u,v)$ is equipped with two pointers $\ell_{uv}$ and $\ell_{vu}$, that reference the faces of $\mathcal E$ lying to the left of $(u,v)$, when traversing such an edge from $u$ to $v$ and from $v$ to $u$, respectively.
\end{itemize}

\begin{figure}[tb!]
  \centering
  \subfigure[Graph $H$ and face $f^*$]{\label{fi:algorithm-example-a}\includegraphics[page=1,height=0.3\textwidth]{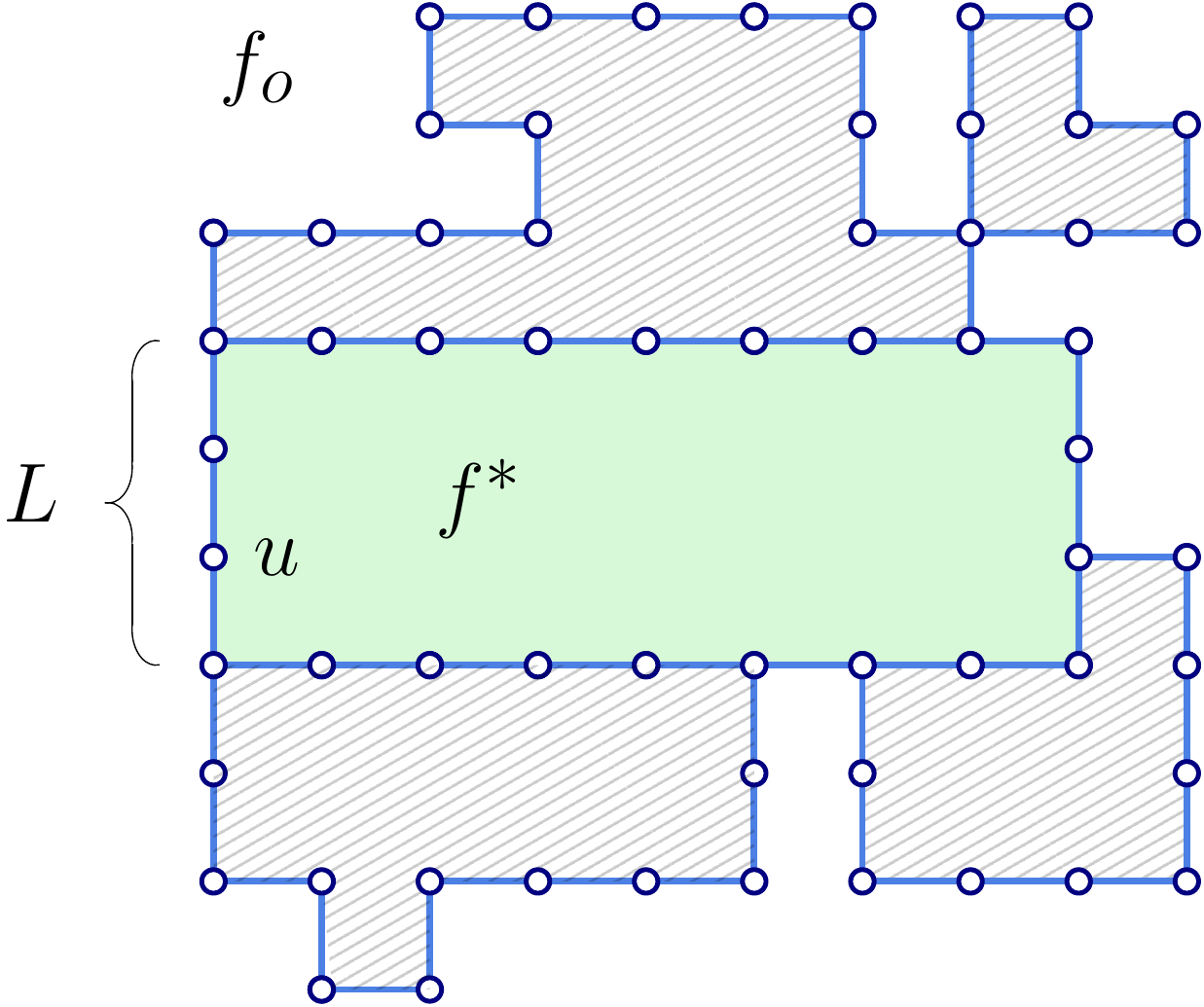}}
  \hfil
  \subfigure[Merging $f^*$ with $f_o$]{\label{fi:algorithm-example-b}\includegraphics[page=2,height=0.3\textwidth]{fig/algorithm-example}}
  \hfil
  \caption{A step of the \algorithmGrep.}
  \label{fig:algorithm_linear}
\end{figure}

At each iteration the algorithm performs the following steps; see \cref{fig:algorithm_linear}.
{\bf Retrieve:} It retrieves an internal face $f^*$ with at least one vertex $u$ with minimum $x$-coordinate (i.e., $x_{\textrm{min}}$) among the placed vertices of $H$; such a vertex is incident to the outer face of $H$.
{\bf Draw:} It assigns coordinates to all the vertices incident to~$f^*$ in such a way that~$f^*$ is drawn as a rectangle $R^*$. Note that such a drawing is unique. Indeed, in any embedding-preserving inner-rectangular unit-length drawing of $H$ in which the cycle delimiting the outer face of each biconnected component of $H$ is the one prescribed, the left side of $R^*$ coincides with the maximal path $L$ containing $u$ that is induced by the placed vertices of $f^*$ with $x$-coordinate equal to $x_{\textrm{min}}$.
{\bf Merge:} It merges $f^*$ with~$f_o$ by suitably changing the pointers of every edge incident to $f^*$, and by removing each edge $(u,v)$ incident to $f^*$ with pointers $\ell_{uv} = \ell_{vu} = f_o$, as well as any resulting isolated vertex. Further, it updates $A$ consequently. Note that, after the merge step, the outer face $f_o$ of the new current graph $H$ is again completely drawn.

\subsection{Details of the Retrieve, Draw, and Merge Steps}\label{se:steps}

We now describe each step in detail.
\begin{description}
\item[Retrieve $f^*$.]

We take the first vertex $u$ in the non-empty bucket $A[x_{\textrm{min}}]$.
Since~$u$ has the smallest $x$-coordinate among the placed vertices of $H$, then $u$ is incident to $f_o$.
Furthermore, since the blocks of $H$ are non-trivial, $u$ has degree either two, three or four in $H$.

Consider first the case in which $u$ has degree $4$. 
Since $u$ is a vertex with smallest $x$-coordinate in $H$ and it is incident to $f_o$, its neighbors must be placed with $x$-coordinates greater than or equal to $x_{\textrm{min}}$.
This is not possible since it would imply that two neighbors of $u$ are drawn on the same grid point.
Hence, Condition C\ref{cond:vertex-same-coordinates} holds and the algorithm stops giving a negative result. 

Consider now the case in which $u$ has degree either two or three (refer to \cref{fi:algorithm-example-a}).
Let $f^*$ be any (of the at most two) internal faces of $H$ incident to $u$.
Let $L$ denote the maximal path containing $u$ that is induced by all the placed vertices of $f^*$ with $x$-coordinate $x_{\textrm{min}}$.
Note that the edges of $L$ are incident to $f_o$, and must form the left side of the rectangle $R^*$ representing $f^*$ in the unit-length grid drawing of $H$ with the given drawing of $f_o$.
Moreover, since all the vertices of the outer face of $H$ have $x$-coordinate greater than or equal to $x_{\textrm{min}}$, such side determines the coordinates of all the vertices of~$f^*$ along $R^*$.

\item[Draw $f^*$.]

We traverse the vertices of $f^*$ while assigning the coordinates determined in the previous step to each vertex.
If there is a vertex of $f^*$ for which Condition C\ref{cond:vertex-different-coordinates} holds, we conclude that the instance is negative, and terminate the algorithm.
Otherwise, each newly placed vertex that was assigned the $x$-coordinate $i$ is inserted at the beginning of $A[i]$ (observe that the vertices placed before drawing $f^*$ are already in $A$).

\item[Merge $f^*$ with $f_o$.]

We traverse counter-clockwise $f^*$ and, for each edge $(u,v)$ that is traversed from $u$ to $v$, we set $\ell_{uv}$ to point to $f_o$.
Then, we remove from $H$ each edge $(u,v)$ with $\ell_{vu}=\ell_{uv}=f_o$ as well as all the resulting isolated vertices, if any (see \cref{fi:algorithm-example-b}).
To finish this step we remove from $A$ all the vertices that were removed from $H$, and update $x_{\textrm{min}}$, if necessary.

\end{description}

\smallskip
\noindent The proof of the next theorem exploits the \algorithmGrep.

\begin{theorem}\label{th:grep}
 The \UIRFElong and \URFElong problems are $O(n)$-time solvable for an $n$-vertex plane graph if the drawing of the outer face is~prescribed.    
\end{theorem}

\begin{proof}
First, as noted previously, it suffices to look at a connected plane graph, as distinct connected components can be dealt with independently. Indeed, the relative positions of such components in the prescribed plane embedding force them to be one outside the other, as otherwise the plane graph would not admit any embedding-preserving inner-rectangular drawing. This implies that the  drawing of the outer face is~prescribed for each of such components. 

In order to prove the theorem, we argue about the correctness and running time of the \algorithmGrep.

We start with the correctness.
Consider that, if the algorithm terminates without a failure, then, by construction, (i) each internal face of $G$ has been drawn as a rectangle, (ii) the rotation system of each vertex has been respected, and (iii) the edges incident to vertices of the cycle delimiting the outer face are drawn as line segments leaving such a cycle towards the interior of the prescribed drawing of the outer face.
Thus, by \cref{lem:planarity}, the drawing is planar.
Again by construction, the coordinates of the vertices on the outer face have not been changed and the edges are horizontal or vertical segments of unit length, hence the drawing is a unit-length grid drawing.

Otherwise, if a failure condition is reached, then we prove that~$G$ does not admit any embedding-preserving unit-length grid drawing where each internal face is drawn as a rectangle and the drawing of the outer face is as prescribed.
Assume that the algorithm fails due to Condition C\ref{cond:vertex-different-coordinates}, i.e., the algorithm is forced to assign different coordinates to the same vertex.
Since by \cref{lem:unicity} if the drawing exists it is unique, then the instance does not admit a grid realization with the prescribed properties.
Assume instead that the algorithm fails due to Condition C\ref{cond:vertex-same-coordinates}, i.e., the algorithm is forced to assign the same coordinates to different vertices. 
This would imply that the drawing is not planar, in contradiction with \cref{lem:planarity}.

We finally prove that the \algorithmGrep runs in $O(n)$ time.
The algorithm performs as many iterations as the internal faces of $G$.
At each iteration on a face $f^*$, it performs a number of operations that is linear in the number of vertices and edges of $f^*$.
Hence, each edge is processed constant number of times, and each vertex is considered at most as many times as the number of incident faces, i.e., at most four times.%
\end{proof}

\cref{th:grep} contrasts with the \NP-hardness results of \cref{th:biconnected,th:biconnected_uir}, where the drawing of the outer face is not prescribed. By again exploiting the observation that any unit-length grid drawing of a cycle with $4$ or $6$ vertices is a rectangle, \cref{th:grep} implies the following result.

\begin{corollary}\label{cor:grep-6}
 Deciding whether a biconnected plane graph admits a unit-length grid drawing is a linear-time solvable problem if the drawing of the outer face is prescribed and all the internal faces have maximum degree~$6$.
\end{corollary}

\section{Algorithms for the URFE and UR problems}\label{se:unit-length-rectangular-drawings}

In this section we study the {\sc Unit-length Rectangular} problem. Since rectangular drawings are convex, the input graphs for  the \UR problem must be biconnected~\cite{DBLP:journals/iandc/BattistaTV01}.

\paragraph{\bf Fixed Embedding.} We start by considering instances with  either a prescribed plane embedding (\cref{th:rect-grep-outer-face}) or  a prescribed planar embedding (\cref{th:rect-grep}). 

\begin{figure}[tb!]
\hfil
\subfigure[Double]{
\includegraphics[width=0.135\textwidth,page=1]{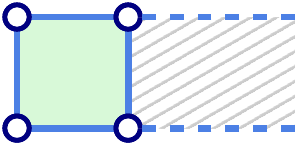}\label{fig:double-corner-face}
}
\hfil
\subfigure[Slim double]{
\includegraphics[width=0.18\textwidth,page=2]{fig/corner-faces.pdf}\label{fig:slim-double-corner-face}
}
\hfil
\subfigure[Fat~double]{
~~\includegraphics[width=0.11\textwidth,page=3]{fig/corner-faces.pdf}~~\label{fig:fat-double-corner-face}
}
\hfil
\subfigure[Degree-4]{
~~\includegraphics[width=0.11\textwidth,page=4]{fig/corner-faces.pdf}~~\label{fig:4-corner-face}
}
\hfil
\subfigure[Degree-6]{
\includegraphics[width=0.15\textwidth,page=5]{fig/corner-faces.pdf}\label{fig:6-corner-face}
}
\hfil
\caption{Corner faces for the proof of \cref{th:rect-grep-outer-face}.}
\label{fig:corner-faces}
\end{figure}

\begin{theorem}\label{th:rect-grep-outer-face}The \URFElong problem is cubic-time solvable for a plane graph $G$ and it is linear-time solvable if all internal faces of $G$ have maximum degree~$6$. 
\end{theorem}
\begin{proof}
  If the input is not biconnected, then we can determine that the instance is negative in linear time~\cite{DBLP:journals/siamcomp/Tarjan72}.
  Hence, in the following, we assume that the input is biconnected, which implies that any face is bounded by a cycle.
  
  In order to solve the \URFE problem in polynomial time, we guess all the possible rectangular grid drawings of the outer face $f_o$.
  For each of them we invoke \cref{th:grep}.
  We have that the rectangular grid drawings of $f_o$ are in one-to-one correspondence (up to a rotation of $90^\circ$, $180^\circ$, or $270^\circ$)  with the possible choices of two vertices that become consecutive corners of the drawing. This corresponds to $O(n^2)$ choices for the drawing of $f_o$.
  For each choice the algorithm \algorithmGrep finds a unit-length grid rectangular drawing in $O(n)$ time, if it exists.
  
  Assume now that all internal faces have maximum degree~$6$.
  Our strategy is to efficiently determine the drawing of the outer face of the input graph $G$ and then to invoke \cref{th:grep} to conclude the proof.

  Note that, if $G$ is a $4$-cycle or a $6$-cycle, then the instance is trivially positive.
  We henceforth assume this is not the case.
  We have also the following simple cases.
  \begin{itemize}
  \item A \emph{double corner face} is a degree-$4$ face with three edges incident to $f_o$, see \cref{fig:double-corner-face}.
  \item A \emph{slim double corner face} is a degree-$6$ face with five edges incident to $f_o$, see \cref{fig:slim-double-corner-face}.
  \item A \emph{fat double corner face} is a degree-$6$ face with four edges incident to $f_o$, see \cref{fig:fat-double-corner-face}.
  \end{itemize}
 If $G$ has a double, slim double, or fat double corner face, then such a face must provide two consecutive $270^\circ$ angles incident any realization of $f_0$ as a rectangle, hence the drawing of the outer face is prescribed and \algorithmGrep can be invoked.
  
  Suppose now that none of the aforementioned cases holds. 
  A \emph{corner face} is a degree-$4$ face that has two edges incident to the outer face $f_o$, see \cref{fig:4-corner-face}, or a degree-6 face that has three edges incident to $f_o$, see \cref{fig:6-corner-face}.
  Observe that, by the assumption that there is no double, slim double, or fat double corner face, a face is incident to a corner of a rectangular drawing of $f_o$ if and only if it is a corner face.
  Hence, there must be exactly four corner faces in order for a rectangular drawing of the input instance to exist, otherwise the input instance is negative.
  The four corner faces can be trivially found in $O(n)$ time.
  They determine a constant number of possible drawings of the outer face as follows.
  If a corner face has degree-$4$, then its degree-$2$ vertex must be a corner of the drawing of the outer face.
  If a corner face has instead degree-$6$, then one of its two degree-$2$ vertices must be a corner of the drawing of the external face.
  Hence we have at most $2^4 = O(1)$ different possible choices for the drawing of the outer face.
  We solve the \URFE problem in this setting by invoking \algorithmGrep with each choice as the prescribed drawing of the outer~face~of~$G$.%
  \end{proof}

By showing that any planar embedding has a unique candidate outer face supporting a unit-length rectangular drawing, we get the following.

\begin{theorem}\label{th:rect-grep}
The \URFElong problem is cubic-time solvable for a planar embedded graph $G$, and it is linear-time solvable if all but at most one face of~$G$ have maximum degree $6$. 
\end{theorem}

\begin{proof}
Observe that, given two rectangles $R_1$ and $R_2$, a necessary condition for drawing $R_2$ inside $R_1$ is that the perimeter of $R_2$ is smaller than the perimeter of $R_1$. 
Hence, given a connected planar embedded graph $G$, we first compute the faces of $G$ with the maximum number of edges in linear time. 
Suppose that there exists exactly one face $f_o$ with the maximum number of edges.
We invoke \cref{th:rect-grep-outer-face} for checking in cubic time (linear, if all the faces different from $f_o$ have degree $6$) if the plane graph consisting of $G$ with the prescribed outer face $f_o$ is a positive or negative instance of \URFE.
Suppose now that there exists more than one face with the maximum number of edges. If $G$ is just an even-length simple cycle, then we conclude that $G$ is a positive instance of \URFE. Otherwise, we conclude the opposite.%
\end{proof}

\paragraph{\bf Variable Embedding.}
Now, we turn our attention to instances with a variable embedding.
We start by providing some relevant properties of the graphs that admit a rectangular (not necessarily unit-length or grid) drawing. 
Let $G$ be one such graph.
To avoid degenerate cases, in what follows, we assume that $G$ is not a cycle, a special case which can be dealt with separately. Let $\Gamma$ be a rectangular drawing of $G$ and let $\Gamma_o$ be the rectangle delimiting the outer face of $\Gamma$.
Refer to \cref{fig:structural}.
Consider the plane graph $G_\Gamma$ corresponding to $\Gamma$.
Since $\Gamma$ is convex, then $G_\Gamma$ is a subdivision of an \emph{internally triconnected} plane graph~\cite[Theorem 1]{DBLP:conf/compgeom/AngeliniLFLPR15}.
That is, every separation pair $\{u, v\}$ of $G_\Gamma$ is such that $u$ and $v$ are incident to the outer face and each connected component of $G_\Gamma \setminus \{u, v\}$ contains a vertex incident to the outer face.

Consider a separation pair $\{u, v\}$ of $G$.
In the following, we provide several useful properties related to $\{u, v\}$. 

\begin{property}\label{prop:internal-vertex}
If at least one of $u$ and $v$ is not in $\Gamma_o$, then there exist exactly two components of $G$ with respect to $\{u,v\}$, one of which is a simple path. Also, the vertices of such a path are drawn on a straight line. See, e.g., the vertices $x_1$ and $y_1$ in \cref{fig:structural}.
\end{property} 
\begin{proof}
The first part of the statement is a consequence of the fact that $G$ is a subdivision of an internally-triconnected plane graph. The second part, instead, follows immediately from the fact that $\Gamma$ is rectangular.%
\end{proof}

\begin{property}\label{pr:three}
If both $u$ and $v$ are in $\Gamma_o$, then there exist either two or three components of $G$ with respect to $\{u,v\}$.
\end{property}
\begin{proof}
The statement follows from the fact that, since $u$ and $v$ are in $\Gamma_o$ and since $\Gamma_o$ is drawn as a rectangle, their degree is at most $3$.%
\end{proof}

\begin{property}\label{pr:opposite}
If both $u$ and $v$ are in $\Gamma_o$ and $G$ has three components $G_1$, $G_2$, and $G_3$ with respect to $\{u,v\}$, then there is exactly one component, say $G_2$, such that $G_2 \setminus \{u,v\}$ does not contain vertices in $\Gamma_o$. Also, $G_2$ is a simple path whose vertices are drawn on a straight line. Furthermore, $u$ and $v$ are drawn on opposite sides of $\Gamma_o$.
Finally, we have that each of $u$ and $v$ has degree $1$ both~in~$G_1$~and in~$G_3$. See, e.g., the vertices $x_2$ and $y_2$ in \cref{fig:structural}.
\end{property}
\begin{proof}
The component $G_2$ must be a simple path, since $G$ is a subdivision of an internally-triconnected plane graph. Also, the vertices of such a path must be drawn either along a horizontal or a vertical line, as otherwise $\Gamma$ would not be rectangular. Finally, since $u$ and $v$ are incident to the outer face and since they both have degree $1$ in $G_2$, we have that each of $u$ and $v$ has degree $1$ both in~$G_1$~and in~$G_3$.%
\end{proof}

\begin{property}\label{prop:no-cross-pair}
There exist no two separation pairs $\{u_1,v_1\}$ and $\{u_2,v_2\}$ of $G$ such that $u_1$ and $v_1$ lie on opposite sides of $\Gamma_o$, $u_2$ and $v_2$ lie on the opposite side of $\Gamma_o$, and $u_1$ and $u_2$ lie on perpendicular sides of $\Gamma_o$.
\end{property}

\begin{proof}
Suppose for a contradiction that there exist two separation pairs $\{u_1,v_1\}$ and $\{u_2,v_2\}$ of $G$ with the properties in the statement. Then there exists an internal face $f_1$ of $\Gamma$ incident to $u_1$ and to $v_1$, and an internal face $f_2$ of $\Gamma$ incident to $u_2$ and to $v_2$. However,
since $\Gamma$ is rectangular, this is possible only if $f_1 = f_2$, which contradicts the assumption that $G$ is not a cycle.%
\end{proof}

\begin{property}\label{prop:opposite-general}
If both $u$ and $v$ are in $\Gamma_o$ and $G$ has two components $G_1$ and $G_2$ with respect to $\{u,v\}$ such that (i) both $G_1$ and $G_2$ are not simple paths in $G$, and (ii) both $u$ and $v$ have degree $2$ in $G_1$, then $u$ and $v$ are drawn on opposite sides of $\Gamma_o$ and $G_1$ contains a path $P_1$ between $u$ and $v$, whose vertices are on a straight line, that is incident to an internal face of $\Gamma$. See, e.g., the vertices $x_3$ and $y_3$ in \cref{fig:structural}.
\end{property}

\begin{proof}
Note that each of $u$ and $v$ is incident to an internal edge that belongs to $G_1$ (possibly the edge $(u,v)$) and each such edge must be incident to the same internal face $f$ of $\Gamma$. Since $f$ is rectangular, the vertices of the subpath $P_1$ of $f$ connecting $u$ and $v$ and passing through these edges must be drawn along a straight line. To complete the proof, we observe that this implies that $u$ and $v$ must be drawn on opposite sides of $\Gamma_o$.%
\end{proof}

The next properties follow directly from the fact that $\Gamma$ is rectangular.

\begin{property}\label{prop:two-same-side}
Suppose that $G$ has two components $G_1$ and $G_2$ with respect to $\{u,v\}$. If $u$ and $v$ are on the same side of $\Gamma_o$, then exactly one of $G_1$ and $G_2$ is a path whose vertices lie in $\Gamma_o$ on a straight line. See, e.g., the vertices $x_4$ and $y_4$ in \cref{fig:structural}.
\end{property}

\begin{property}\label{prop:two-perpendicular-sides}
Suppose that $G$ has two components $G_1$ and $G_2$ with respect to $\{u,v\}$.
If $u$ and $v$ are incident to perpendicular sides of $\Gamma_o$, then exactly one of $G_1$ and $G_2$, say $G_1$, is a simple path. Moreover, $G_1$ is drawn in $\Gamma$ as an orthogonal polygonal line with a single bend. See, e.g., the vertices $x_5$ and $y_5$ in \cref{fig:structural}.
\end{property}

\begin{property}\label{prop:series}
Suppose that $G$ has two components $G_1$ and $G_2$ with respect to $\{u,v\}$.
If $u$ and $v$ are on opposite sides of $\Gamma_o$, then each of $u$ and $v$ has degree $1$ in at least one of $G_1$ and $G_2$. See, e.g., the vertices $x_6$ and $y_6$ in \cref{fig:structural}.
\end{property}

A \emph{caterpillar} is a tree such that removing its leaves results in a path, called \emph{spine}, possibly composed of a single node.
The \emph{pruned SPQR-tree} of a biconnected planar graph $G$, denoted by 
$T^*$, is the tree obtained
from the SPQR-tree $T$ of $G$, after removing the $Q$-nodes~of~$T$.

\begin{lemma}\label{lem:structural}
Let $G$ be a graph that admits a rectangular drawing.
Then the pruned SPQR-tree $T^*$ of $G$ is a caterpillar with the following properties:
\begin{enumerate}[\bf (i)]
\item\label{cond:leaves-S}All its leaves are $S$-nodes;
\item\label{cond:no-adjacent-R-and-R}its spine contains no two adjacent $R$-nodes;
\item\label{cond:no-adjacent-R-and-P}its spine contains no two adjacent nodes $\mu$ and $\nu$ such that $\mu$ is a $P$-node and $\nu$ is an $R$-node;  
\item\label{cond:P-three} each $P$-node $\mu$ has exactly $3$ neighbors in $T$;
and
\item\label{cond:$S$-node-skeleton}the skeleton of each $S$-node of the spine of $T^*$ contains two chains of virtual edges corresponding to $Q$-nodes, separated by two virtual edges, each corresponding to either a $P$- or an $R$-node.
\end{enumerate}
\end{lemma}

\begin{proof}
\begin{figure}[tb!]
\includegraphics[width=\textwidth,page=6]{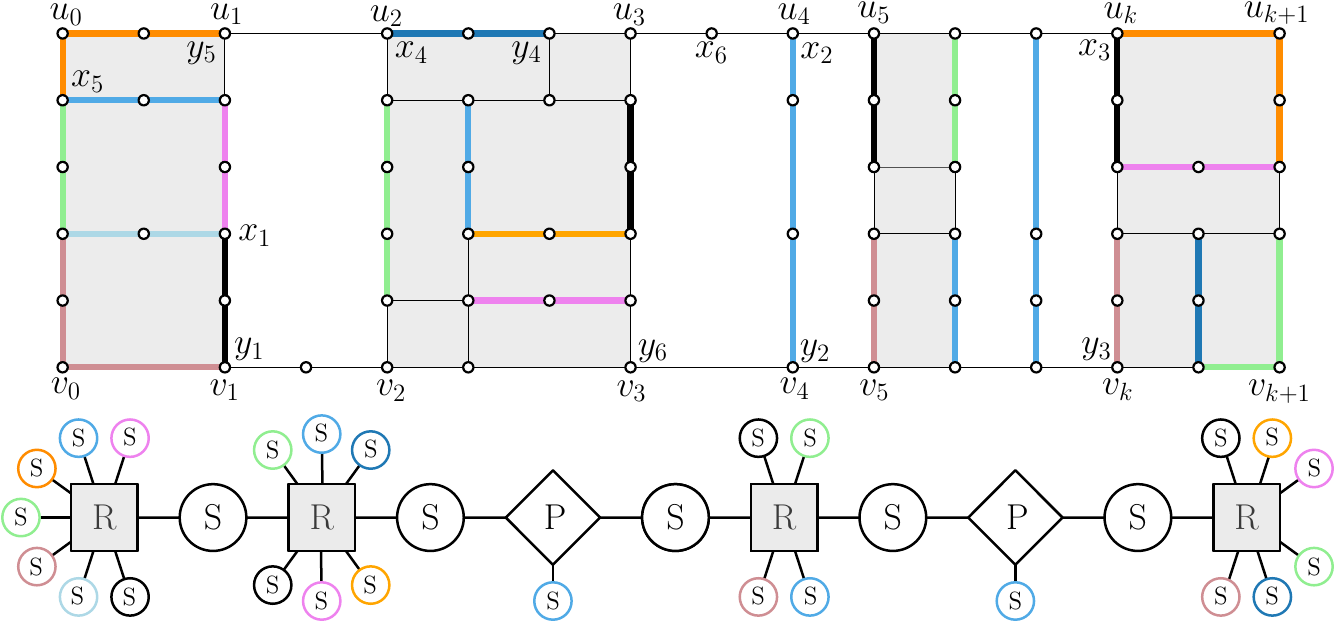}
\caption{
A rectangular grid drawing of a planar graph and its pruned SPQR-tree $T^*$. $S$-, $P$-, and $R$-nodes are circles, rhombuses and squares, respectively. The subgraphs corresponding to $S$-nodes that are leaves of $T^*$ are thick.}
\label{fig:structural}
\end{figure}
In the following, we assume that $G$ is not a cycle, as otherwise the statement trivially holds. With this assumption, $T^*$ contains at least one $P$- or $R$-node.

Let $\Gamma$ be a rectangular drawing of $G$, and let $\Gamma_o$ be the drawing of the outer face of $\Gamma$. Refer to \cref{fig:structural}.

Suppose, for a contradiction, that there exist two adjacent $R$-nodes $\mu$ and $\nu$ in the spine of $T^*$. Let $\{u,v\}$ be the separation pair shared by their skeletons, and let $e_{\mu,\nu}$ and $e_{\nu,\mu}$ be the virtual edges in $\skel(\nu)$ and in $\skel(\mu)$ corresponding to $\mu$ and to $\nu$, respectively.
By \cref{prop:internal-vertex}, both $u$ and $v$ must lie in $\Gamma_o$. 
By \cref{prop:two-same-side,prop:two-perpendicular-sides}, $u$ and $v$ must lie on opposite sides of $\Gamma_o$.
Therefore, by \cref{prop:series}, each of $u$ and $v$ has degree $1$ in at least one of $\expan{(e_{\mu,\nu})}$ and $\expan{(e_{\nu,\mu})}$, which implies that either $\mu$ or $\nu$ is an $S$-node. Therefore, we get a contradiction. This proves {\bf Condition (\ref{cond:no-adjacent-R-and-R})} of the statement.

By \cref{prop:internal-vertex}, the poles of a $P$-node of $T$ are incident to $\Gamma_o$. By \cref{pr:opposite}, the neighbors of a $P$-node are either $S$- or $Q$-nodes. This proves {\bf Condition (\ref{cond:no-adjacent-R-and-P})}.

A  $P$-node $\mu$ of $T$ has at least three neighbors in $T$, by definition. By \cref{prop:internal-vertex,pr:three}, we have that any node of $T$ has at most three neighbors in $T$. This proves {\bf Condition (\ref{cond:P-three})}.

Next, we show that $T^*$ is a caterpillar, and that it satisfies {\bf{Conditions (\ref{cond:leaves-S})}} and {\bf (\ref{cond:$S$-node-skeleton})} of the statement. We distinguish two cases.

\emph{Case 1: there exists no separation pair $\{u,v\}$ of $G$ such that $u$ and $v$ are on opposite sides of $\Gamma_o$.} 
In this case, by \cref{pr:opposite}, $T^*$ contains no $P$-nodes. For any separation pair $\{u, v\}$ of $G$, we have that  (i) there exist exactly two components of $G$ with respect to $\{u, v\}$ and that (ii) exactly one of the components of $G$ with respect to $\{u, v\}$ corresponds to an $S$-node, which is a simple path. This comes from \cref{prop:internal-vertex} if at least one of $u$ and $v$ is an internal vertex, from \cref{prop:two-perpendicular-sides} if $u$ and $v$ lie on perpendicular sides of $\Gamma_o$, and from \cref{prop:two-same-side} if $u$ and $v$ lie on the same side of $\Gamma_o$. It follows that there exist no two $R$-nodes in $T$. Note that $T$, and hence $T^*$, contains $S$-nodes, as the corners of $\Gamma_0$ are vertices of $G$ with degree $2$. Therefore, $T^*$ is a star whose leaves are $S$-nodes and whose central vertex is an $R$-node. This proves {\bf{Condition (\ref{cond:leaves-S})}}; also, {\bf Condition (\ref{cond:$S$-node-skeleton})} is vacuously true.

\emph{Case 2: There exists a separation pair $\{u,v\}$ of $G$ such that $u$ and~$v$ are on opposite sides of $\Gamma_o$.} By \cref{prop:no-cross-pair}, any other separation pair $\{u',v'\}$ different from $\{u,v\}$ where $u'$ and $v'$ are on opposite sides of $\Gamma_o$ is such that either $u$ and $u'$ are on the same side of $\Gamma_o$ or $u$ and $v'$ are on the same side~of~$\Gamma_o$. 
Therefore, after a possible rotation by a multiple of $90^\circ$, in the following we will assume that $u$ lies on the top side of $\Gamma_o$ and $v$ lies on the bottom side of~$\Gamma_o$.
Let $S = [\{u_1,v_1\},\{u_2,v_2\},\dots,\{u_k,v_k\}]$ be the separation pairs of $G$ such that $u_1, u_2,\dots, u_k$ lie, in this left-to-right order, on the top side of $\Gamma_o$, such that $v_1, v_2,\dots, v_k$ lie, in this left-to-right order,  on the bottom side of $\Gamma_o$, such that all these vertices have degree $3$, and such that $u_i$ and $v_i$ share the same $x$-coordinate, for $i=1,\dots,k$.
The next claim shows~that~$k\geq 1$. 

\begin{claimx}\label{cl:good-pairs}
If there exists a separation pair $\{u,v\}$ of $G$ such that $u$ and $v$ are on opposite sides of $\Gamma_o$, then 
there exists at least one separation pair $\{u',v'\}$ such that:
\begin{itemize}
\item $u'$ and $v'$ are on opposite sides of $\Gamma_o$;
\item $u'$ and $v'$ share the same $x$-coordinate in $\Gamma$;  
\item $u'$ and $v'$ have degree $3$ in $G$; and
\item there exists a path  between  $u'$ and $v'$ drawn along a vertical line that is incident to an internal~face~of~$\Gamma$.
\end{itemize}

\end{claimx}
\begin{claimproof}
The proof distinguishes two cases.

\emph{Suppose first that at least one of $u$ and $v$, say $v$, has degree $3$.} We show that there exists a vertex $u'$ lying along the top side of $\Gamma_o$, where possibly $u' =u$, such that the separation pair $\{u',v\}$ satisfies the properties required by the claim.
Note that, since $v$ has degree $3$ and is incident to the bottom side of $\Gamma_o$, two of its neighbors lie along the bottom side of $\Gamma_o$. Therefore, the third neighbor of $v$, must either be a vertex $u^*$ incident to the top side of $\Gamma_o$ (possibly $u^* = u$) or an internal vertex $i_v$ lying vertically above $v$. In the former case, since $\Gamma$ is rectangular, we have that $u^*$ has degree $3$, lies vertically above $v$ in $\Gamma$ (which implies that $u^*$ and $v$ have the same $x$-coordinate), and is connected to $v$ via a path (indeed, a single edge) drawn along a vertical line. Thus, setting $u'=u^*$ yields the desired separation pair. 
In the latter case,  since $\{u,v\}$ is a separation pair, there exists an internal face $f$ shared by $u$, $v$, and $i_v$. Let $u'$ be the first vertex of the top side of $\Gamma_o$ that is encountered when traversing the boundary of $f$ starting at $v$ and passing through $i_v$. Consider the subpath of $f$ between $v$ and $u'$ that contains $i_v$.
Since $i_v$ lies vertically above $v$ in $\Gamma$, and since $\Gamma$ is rectangular, this path must be drawn as a straight-line segment between $v$ and $u'$, which implies that $u'$ has degree $3$ and has the same $x$-coordinate as $v$. Therefore, since both $u'$ and $v$ belong to $f$ and lie on opposite sides of $\Gamma_o$, they form the sought separation pair.

\emph{Suppose next that both $u$ and $v$ have degree-$2$.} Consider the internal face $f$ of~$\Gamma$
shared by $u$ and $v$. We show that there exists a separation pair that satisfies the properties of the statement whose vertices are incident to $f$. 
If $f$ contains no degree-$3$ vertex incident to the top side of $\Gamma_o$ and no degree-$3$ vertex incident to the bottom side of $\Gamma_o$, then both the paths of $G$ that form the top and the bottom side of $\Gamma_o$ belong to $f$, hence~$G$ is a cycle, which contradicts the assumptions of the lemma. Otherwise, $f$ contains a degree-$3$ vertex $u'$ incident to the top side of $\Gamma_o$ and a vertex $v'$ incident to the bottom side of $\Gamma_o$ (or a degree-$3$ vertex $v'$ incident to the bottom side of $\Gamma_o$ and a vertex $u'$ incident to the top side of $\Gamma_o$). Then the existence of the sought separation pair can be deduced as in the first case of the proof, with $u'$ and $v'$ playing the role of $u$ and $v$. 
\end{claimproof}

\noindent
We set $L = \{u_0,v_0\} \circ S \circ \{u_{k+1},v_{k+1}\}$, where $u_0$, $v_0$, $u_{k+1}$, and $v_{k+1}$ are the vertices of $G$ lying at the top-left, bottom-left, top-right, and bottom-right corner of $\Gamma_o$, respectively, where $\circ$ denotes the concatenation operator. For $i=0,\dots,k+1$, let $P_i$ be the path in $G$ connecting $u_i$ and $v_i$ that is drawn along a vertical line in $\Gamma$. This path exists by the previous claim, for $i=1,\dots,k$, and since $\Gamma_o$ is a rectangle, for $i\in \{0,k+1\}$. Also, let $C_i$ be the cycle of $G$ that contains $u_i$, $u_{i+1}$, $v_{i+1}$, and $v_{i}$, that contains $P_i$ and $P_{i+1}$, and that is drawn as a rectangle in $\Gamma$. Clearly, any two cycles $C_i$ and $C_{i+1}$ share the path $P_{i+1}$.
We denote by $G_i$ the subgraph of $G$ induced by the vertices in the interior and along the boundary of~$C_i$. 

We show that $T^*$ can be constructed iteratively starting from the empty tree, as follows. At each point of the construction, $T^*$ will be a caterpillar whose spine does not have a $P$-node as an end-point. Also, a leaf of $T^*$  will be denoted as \emph{active} and will be used in the subsequent iteration, if any, as an attachment endpoint to extend $T^*$.

The construction of $T^*$ starts by considering the following two cases.
\begin{itemize}
    \item If $G_0 = C_0$, we introduce an $S$-node $\mu_0$ in $T^*$. In particular, $(u_1,v_1)$ is a virtual edge of $\skel(\mu_0)$, and the other virtual edges of $\skel(\mu_0)$ correspond to the edges of $C_0$ incident to $\Gamma_0$.
\item  Otherwise, $G_0 \neq C_0$. Consider a separation pair $\{u,v\}$ of $G_0$.
By \cref{cl:good-pairs} applied to $G_0$ and since $\{u_1,v_1\}$ is the first pair in $S$,  we have that $u$ and $v$ do not lie one on the top and one on the bottom side of $\Gamma$. Therefore, by \cref{prop:internal-vertex,prop:two-same-side,prop:two-perpendicular-sides}, one of the two components of $G_0$ with respect to $\{u,v\}$ is a simple path. Thus, $G_0$ is the subdivision of a triconnected planar graph. 
Hence, we introduce an $R$-node $\mu_0$ in $T^*$ whose skeleton is obtained by replacing each maximal induced path in $G_0$ not containing $u_1$ or $v_1$ in its interior  with a virtual edge. For each of such virtual edges that does not correspond to a single real edge, we add to $T^*$ an $S$-node adjacent to $\mu_0$; note that there are at least two $S$-nodes adjacent to $\mu_0$, as $u_0$ and $v_0$ have degree $2$ in $G$. We also introduce in $\skel(\mu_0)$ the virtual edge $(u_1,v_1)$. 
\end{itemize}
In both cases (i.e., $G_0 = C_0$ and $G_0 \neq C_0$), $\mu_0$ is the active endpoint of $T^*$.

Next, for $i=1,\dots,k$, we consider the separation pair $\{u_i,v_i\}$.
Denote by $\xi$ the active endpoint of the spine (right before considering the current index $i$). Then $\xi$ is either an $S$-node or an $R$-node. Also, $\skel(\xi)$ contains a virtual edge $(u_i,v_i)$; note that this is true for $i=1$, as described above. As before, we distinguish two cases.

\begin{itemize}
\item \emph{Suppose that $G_i = C_i$.} We have two further cases.
\begin{itemize}
\item If $\xi$ is an $S$-node, then we introduce a $P$-node $\mu_{i,1}$ in $T^*$ adjacent to $\xi$ and either one or two more $S$-nodes adjacent to $\mu_{i,1}$. In particular, the skeleton of $\mu_{i,1}$ (in $T$) is a bundle of three parallel edges $(u_i,v_i)$. If $P_i$ is a single edge, then we add to $T^*$ a single $S$-node $\mu_{i,3}$ adjacent to $\mu_{i,1}$, while if $P_i$ is not a single edge, then we add to $T^*$ two more $S$-nodes $\mu_{i,2}$ and $\mu_{i,3}$ adjacent to $\mu_{i,1}$. If $P_i$ is not a single edge, then the skeleton of $\mu_{i,2}$ is a cycle containing one virtual edge for each edge of the path $P_i$ plus a virtual edge $(u_i,v_i)$. The skeleton of $\mu_{i,3}$ is a cycle consisting of a virtual edge $(u_i,v_i)$, followed by a virtual edge for each horizontal edge in the top side of $C_i$, followed by one virtual edge $(u_{i+1},v_{i+1})$, followed by a virtual edge for each horizontal edge in the bottom side of $C_i$. Finally, we set $\mu_{1,3}$ as the active node of $T^*$.
\item If $\xi$ is an $R$-node, then we introduce an $S$-node $\mu_i$ in $T^*$ adjacent to $\xi$ whose skeleton is a cycle consisting of a virtual edge $(u_i,v_i)$, followed by one virtual edge for each horizontal edge in the top side of $C_i$, followed by a path $P^*$ of virtual edges defined below, followed by one virtual edge for each horizontal edge in the bottom side of $C_i$.
If $i < k$, then the path $P^*$ consists of the single virtual edge $(u_{i+1},v_{i+1})$; otherwise, if $i = k$, then the path $P^*$ contains a virtual edge for each real edge incident to the right side of $\Gamma_o$ (i.e., for each edge of the right side of $C_k$). 
Finally, we set $\mu_{i}$ as the active endpoint of $T^*$.
\end{itemize}
\item \emph{Suppose now that $G_i \neq C_i$.}
With the same motivation as for $G_0$, we introduce an $R$-node $\mu_i$ in $T^*$ adjacent to $\xi$ whose skeleton is obtained by replacing each maximal induced path in $G_i$ that does not contain $u_i$, $v_i$, $u_{i+1}$, or $v_{i+1}$ with a virtual edge. We add an $S$-node for each of such virtual edges that does not correspond to a single real edge. Also, we introduce in $\skel(\mu_i)$ the virtual edge $(u_i,v_i)$ and, unless $i=k$, the virtual edge $(u_{i+1},v_{i+1})$. Finally, we set $\mu_i$ as the active endpoint of $T^*$.
\end{itemize}

It is easy to observe that, after each step of the inductive construction of $T^*$, the tree $T^*$ is a caterpillar whose spine connects $\mu_0$ with the active endpoint of $T^*$, hence $T^*$ is eventually a caterpillar. Also, the construction guarantees that {\bf{Condition (\ref{cond:leaves-S})}} and {\bf{Condition (\ref{cond:$S$-node-skeleton})}} of the statement are satisfied.%
\end{proof}

\begin{figure}[tb!]
\centering
\includegraphics[scale=.6,page=5]{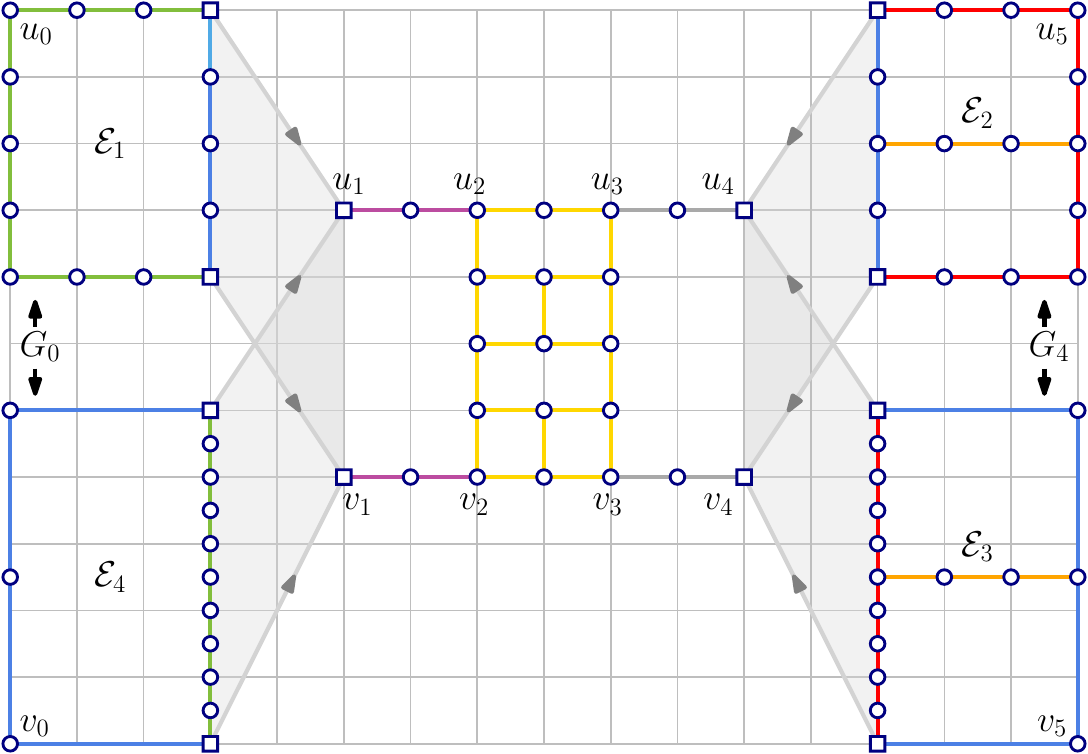}

\caption{
Four plane embeddings of a graph $G$ that support a rectangular drawing of $G$, obtained by selecting one of the plane embeddings ${\cal E}_1$ and ${\cal E}_4$ of the subgraph $G_0$ of $G$ and one of the the plane embeddings
${\cal E}_2$ and ${\cal E}_3$ of the subgraph $G_4$~of~$G$. Only the embeddings ${\cal E}_1$ and ${\cal E}_2$ support a unit-length rectangular drawing.
}
\label{fig:four-embeddings}
\end{figure}

Consider a graph $G$ that satisfies the conditions of \cref{lem:structural}. If the spine of the pruned SPQR-tree of $G$ contains at least two nodes or at least one $P$-node, we say that $G$ is \emph{flat}; otherwise, $G$ is the subdivision of a triconnected planar graph.
Exploiting \cref{lem:structural}, we can prove the following; refer to \cref{fig:four-embeddings}.

\begin{lemma}\label{lem:constant-embeddings}
Let $G$ be an $n$-vertex graph. The following hold: 
\begin{itemize}
\item
All the unit-length rectangular drawings of $G$, if any, have the same plane embedding~$\cal E$ (up to a reflection), which can be computed in $O(n)$ time.

\item 
If $G$ is flat, all the rectangular drawings of $G$, if any, have at most four possible plane embeddings (up to a reflection), which can be computed in~$O(n)$~time.
\end{itemize}
\end{lemma}

\begin{proof}
We prove the first part of the statement.

Suppose that $G$ has a unit-length rectangular drawing $\Gamma$. By \cref{lem:structural}, the pruned SPQR-tree $T^*$ of $G$ is a caterpillar. By 
{\bf{Condition (\ref{cond:leaves-S})}} of \cref{lem:structural}, all the nodes of $T^*$ that are not in the spine are $S$-nodes, hence all the planar embeddings of $G$ are obtained by embedding the skeletons of the $P$- and $R$-nodes of the spine of $T^*$. 

We arbitrarily select a normal orientation $T^*$ such that its spine is a directed path, and visit the spine $\mu_1,\dots,\mu_k$ of $T^*$ according to such an  orientation. 
Note that neither $\mu_1$ nor $\mu_k$ can be an $S$-node, by 
{\bf{Condition (\ref{cond:leaves-S})}} of \cref{lem:structural} and since $T$ does not contain any two adjacent $S$-nodes. We construct the plane embedding $\cal E$ of $G$ and select its outer face $f_o$ as follows. All the choices we perform are obliged, as a consequence of \cref{lem:structural} and of \cref{prop:internal-vertex,pr:three,pr:opposite,prop:no-cross-pair,prop:opposite-general,prop:two-same-side,prop:two-perpendicular-sides,prop:series}.

\emph{Suppose that $\mu_1$ is a $P$-node.} By \cref{prop:internal-vertex}, the poles of $\mu_1$ are incident to $f_o$. By \cref{pr:opposite}, we have that either (i) exactly one neighbor $\nu$ of $\mu_1$ is a $Q$-node, or (ii) exactly one neighbor $\nu$ of $\mu_1$ is an $S$-node corresponding to a simple path in $G$, or (iii) at least two neighbors of $\mu_1$ are $S$-nodes corresponding to a simple path in $G$. 
In cases (i) and (ii), the virtual edge of $\skel(\mu_1)$ corresponding to $\nu$ must lie in between the other two virtual edges.
In case (iii), since $\Gamma$ is rectangular, one of the simple paths corresponding to the neighbors of $\mu$ must be shorter than the others. The corresponding virtual edge must lie in between the other virtual edges in the embedding of $\skel(\mu_1)$.
Two cases are possible: Either $\mu_1$ is the unique node of the spine of $T^*$ or not.
In the former case, the virtual edges corresponding to the remaining two neighbors of $\mu_1$ in $T^*$ can be ordered arbitrarily. Note that this yields exactly two planar embeddings of $G$ that are one the reflection of the other.
Otherwise, we set the virtual edge corresponding to $\mu_2$ at the rightmost virtual edge in the embedding of $\skel(\mu_1)$. 

\emph{Suppose that $\mu_1$ is an $R$-node.}
Two cases are possible: Either $\mu_1$ is the unique node of the spine of $T^*$ or not.
In the former case, $G$ is the subdivision of a triconnected planar graph. Hence, it has a unique planar embedding $\cal E$, up to a reflection. Since in any unit-length rectangular drawing of $G$, the outer face must be bounded by a face of $\cal E$ of maximum size and since no internal face may have the same size of the outer face, we can determine in $O(n)$ time whether $\cal E$ does not support a rectangular drawing or whether a candidate outer face of $\cal E$ exists.
This determines a unique candidate plane embedding of $G$, up to a reflection.
In the latter case, consider the virtual edge $e_2$ corresponding to $\mu_2$ in $\skel(\mu_1)$. 
Recall that, since $\mu_1$ is a $R$-node, $\skel(\mu_1)$ admits a unique (up to a reflection) planar embedding.
In such an embedding, we remove the edge $e_1 = (u_1,v_1)$, and let $\skel^-(\mu_1)$ be the resulting embedded graph.
Note that, by {\bf Condition (\ref{cond:leaves-S})} of \cref{lem:structural}, each virtual edge of $\skel^-(\mu_1)$ corresponds to a simple path in $G$. Let $G^-$ be the embedded subgraph of $G$ obtained by replacing each virtual edge of $\skel^-(\mu_1)$ with the associated path. Let $P$ and $P_1$ be the two paths of $G^-$ between $u_1$ and $v_1$ that share the same face of $G^-$ (note that, they stem from the face of $\skel^-(\mu_1)$ that used to host the edge $e_1$). Since $\Gamma$ is unit-length and rectangular, one of $P$ and $P_1$, say $P_1$, is shorter than the other. We select the embedding of $\skel(\mu_1)$ so that the path of $\skel(\mu_1)$ that corresponds to $P_1$ is incident to the right outer face of the embedding of $\skel(\mu_1)$. 

Consider now a node $\mu_i$, with $1<i<k$. Let $e_{i-1}$ and let $e_{i}$ be the virtual edges of $\skel(\mu_i)$ corresponding to $\mu_{i-1}$ and $\mu_{i+1}$. If $\mu_i$ is an $S$-node, then there is no embedding choice to perform. Otherwise, we exploit the following observation. Since a rectangular drawing of $G$ is convex, the separation pairs corresponding to the poles of $P$- and $R$-nodes must be incident to the outer face of any rectangular drawing of $G$~\cite{DBLP:journals/iandc/BattistaTV01,DBLP:books/ws/NishizekiR04}. Therefore, the embedding choices for $\mu_i$ are described below.

\emph{Suppose that $\mu_i$ is a $P$-node.}
Let $\nu$ be the neighbor of $\mu_i$ different from $\mu_{i-1}$ and $\mu_{i+1}$.
By \cref{pr:opposite}, we have that either (i) $\nu$ is a Q-node, or (ii) $\nu$ is an $S$-node corresponding to a simple path in $G$.
In both cases, the virtual edge of $\skel(\mu_i)$ corresponding to $\nu$ lies in between $e_{i-1}$ and $e_{i}$. Also, we let $e_{i-1}$ and $e_{i}$ be the leftmost and the rightmost virtual edges in the embedding of $\skel(\mu_i)$, respectively.

\emph{Suppose that $\mu_i$ is an $R$-node.}
Recall that, since $\mu_i$ is an $R$-node,  $\skel(\mu_i)$ admits a unique (up to a reflection) planar embedding. We select the embedding of $\skel(\mu_i)$ so that $e_{i-1}$ and $e_{i}$ are incident to the left outer face and to the right outer face of such an embedding, respectively.

Finally, consider now the node $\mu_k$. The embedding of $\skel(\mu_k)$ can be selected, based on its type, as described for $\mu_1$.

Now, we prove the second part of the statement. Recall that the separation pairs corresponding to the poles of $P$- and $R$-nodes must be incident to the outer face of any rectangular drawing of $G$. Therefore, the embedding choices for the $P$- and $R$-nodes $\mu_i$, with $1 < i < k$, in an embedding that supports a rectangular drawing are obliged and correspond to the ones described above. Also, for the $S$-nodes there are no embedding choices. Therefore, the only remaining embedding choices occur on $\mu_1$ and $\mu_k$.

If $k=1$, then the spine of $T^*$ contains a single node. If $\mu_1 = \mu_k$ is an $R$-node, then $G$ is not flat and it is the subdivision of a triconnected planar graph and there is nothing to prove. Otherwise, $\mu_1 = \mu_k$ is a $P$-node, and $G$ consists of three paths sharing their end-vertices. Therefore, it admits three plane embeddings, up to a reflection.

If $k>1$, consider node $\mu_1$. If $\mu_1$ is an $R$-node, then consider the subgraph $G_0$ of $G$ corresponding to it. Namely, let $e_{\mu_1,\mu_2}$ the virtual edge of $\skel(\mu_2)$ corresponding to $\mu_1$, then $G_0 = \expan(e_{\mu_1,\mu_2})$. As shown above, $G_0$ is the subdivision of a triconnected planar graph. Since it admits a unique planar embedding (up to a reflection), since there exists a unique face of such an embedding that contains the poles of $\mu_1$, and since these vertices must be incident to the outer face of a plane embedding of $G$ that supports a rectangular drawing of $G$, we have that $G_0$ admits only two candidate plane embeddings.
If $\mu_1$ is a $P$-node, then let $\nu_1$, $\nu_2$, and $\nu_3$ be its three neighbors (see \cref{lem:structural}) and let $\nu_3$ be its neighbor in the spine of $T^*$. By \cref{lem:structural},  $\nu_1$ and $\nu_2$ are $S$-nodes whose corresponding subgraph of $G$ is a simple path, whereas the subgraph of $G$ corresponding to $\nu_3$ is not a simple path. 
Note that, because any rectangular drawing is also convex, the embedding $\mathcal{E}_1$ of $\skel(\mu_1)$ induced by any embedding of $G$ that supports a rectangular drawing is such that the virtual edge corresponding to $\nu_3$ is incident to the outer face of $\mathcal{E}_1$.
It follows that the only two possible choices to determine a candidate embedding of $\skel(\mu_1)$ depend on the fact that the virtual edge corresponding to $\nu_1$ is before or after the virtual edge corresponding to $\nu_2$. 
The degrees of freedom of the embeddings of $\skel(\mu_k)$ are analogous. Hence, if $k \neq 1$, we have four possible plane embeddings of $G$, up to a reflection.%
\end{proof}

The next theorem shows that the \UR problem is polynomial-time solvable. Surprisingly, the problem seems to be harder for non-flat instances.

\begin{theorem}\label{th:linear-cubic-dichotomy}
The \URlong problem is cubic-time solvable. Also, if the input planar graph is flat, then the \URlong problem is linear-time solvable.
\end{theorem}

\begin{proof}
First, we test whether $G$ satisfies the conditions of \cref{lem:structural}, which can clearly be done in $O(n)$ time, where $n$ is the number of vertices of $G$, by computing and visiting $T^*$. We reject the instance if this test fails. 

If $G$ is not flat, then the spine of $T^*$ consists of a single $R$-node, by definition, hence $G$ is the subdivision of a triconnected planar graph and it admits a unique planar embedding, up to reflection. Hence, we can test whether $G$ admits a unit-length rectangular drawing in $O(n^3)$ time by means of \cref{th:rect-grep}.

If $G$ is flat, then, by means of \cref{lem:constant-embeddings}, we compute in $O(n)$ time the unique candidate plane embedding $\cal E$ of $G$ that may support a unit-length rectangular drawing of $G$, if any. Let $f_o$ be the outer face of $\cal E$. We show that there exists a unique candidate drawing $\Gamma_o$ of $f_o$, that is, in any unit-length rectangular drawing of $G$ the outer face is delimited by the same rectangle $\Gamma_o$, up to a reflection and and a rotation. Provided this statement, we can use \cref{th:grep} to test in $O(n)$ time for the existence of a unit-length rectangular drawing of $G$ that respects the plane embedding $\cal E$ and such that $f_o$ is drawn as $\Gamma_o$.

We now show that there exists a unique candidate drawing $\Gamma_o$ of $f_o$. We distinguish two cases, depending on whether the spine of $T^*$ contains a $P$-node {\bf (Case 1)} or not ({\bf Case 2}).
In {\bf Case 1}, let $\mu$ be a $P$-node of the spine of $T^*$, and let $u$ and $v$ be the poles of $\mu$. By \cref{pr:opposite}, these vertices lie on opposite sides of the rectangle $\Gamma_o$ bounding the outer face of any rectangular drawing $\Gamma$ of $G$ and there exists exactly one component of $G$ with respect to $\{u,v\}$ that is a simple path $P$ whose vertices are drawn on a straight line in $\Gamma$.
In {\bf Case 2}, let $\mu$ be an $R$-node of the spine of $T^*$, and let $u$ and $v$ be the poles of $\mu$. Since the spine of $T^*$ does not contain any $P$-node and since it is not a single R-node, there exists a neighbor $\nu$ of $\mu$ in the spine of $T^*$ that is an $S$-node. By \cref{prop:opposite-general}, vertices $u$ and $v$ lie on opposite sides of the rectangle $\Gamma_o$ bounding the outer face of any rectangular drawing $\Gamma$ and there exists a path $P$ between $u$ and $v$ that belongs to the component of $G$ with respect to $\{u,v\}$ corresponding to $\mu$; also, the path $P$ is incident to an internal face of $\Gamma$ and its vertices are drawn on a straight line.
Both in {\bf Case 1} and in {\bf Case 2}, let $|P|$ and $|f_o|$ denote the length of $P$ and $f_o$, respectively.
By \cref{pr:opposite}, up to a $90^\circ$ rotation of $\Gamma$, the value $|P|$ must correspond to the height of $\Gamma$, whereas $(|f_o| - 2|P|)/2$ must correspond to the width of $\Gamma$. Note that, if the latter value is less than or equal to zero, then $G$ does not admit any unit-length rectangular drawing, in which case we reject the instance.
Let $r$ (resp. $\ell$) be the number of edges traversed when walking in clockwise (resp. counter-clockwise) direction from $u$ and $v$ along the cycle $C_o$ bounding $f_o$. By the above discussion, the four vertices  $u_r$, $v_r$, $v_\ell$, and $u_\ell$ that lie at the corners of the rectangle $\Gamma_o$ bounding the outer face of any rectangular drawing $\Gamma$ of $G$ are the vertices at distance $(r - |P|)/2$, $(r + |P|)/2$, $r+ (\ell - |P|)/2$ and $r+ (\ell + |P|)/2$ in clockwise direction from $u$ along $C_o$. It follows that $\Gamma_o$ is uniquely defined, which proves the statement and hence the theorem.
\end{proof}

The techniques we developed for unit-length drawings allow us to prove the following.

\begin{theorem}\label{th:non-unit-rectangular-drawing}
The problem of testing for the existence of a rectangular drawing of an $n$-vertex planar graph $G$ is solvable in $O(n^2 \log^3 n)$ time. Also, if $G$ is flat, the problem is solvable in $O(n \log^3 n)$ time.
\end{theorem}

\begin{proof}
First, we test whether $G$ satisfies the conditions of \cref{lem:structural}, which can clearly be done in $O(n)$ time by computing and visiting $T^*$, and reject the instance if this test fails.

We start by considering the case in which $G$ is flat.
Due to \cref{lem:constant-embeddings}, only up to four plane embeddings of $G$ are candidates for a rectangular drawing of $G$ that respects them. Also, such embeddings can be computed in $O(n)$ time. For each of them, we test for the existence of a rectangular drawing respecting it by solving a max-flow problem on a linear-size planar network with multiple sources and sinks in $O(n \log^3 n)$ time~\cite{DBLP:journals/siamcomp/BorradaileKMNW17}. Such a network can be defined following Tamassia's~\cite{DBLP:books/ph/BattistaETT99} classic approach to test for the existence of rectilinear drawings of plane graphs.
In such a network $\cal N$, we have that:
\begin{itemize}
    \item each node of $\cal N$ corresponding to a vertex of $G$ is a source producing $4$ units of flow, each corresponding to a $90^\circ$ angle;
    \item each node of $\cal N$ corresponding to a face $f$ of $G$ is a sink consuming $2|f|-4$ (resp. $2|f|+4$) units of flow if $f$ is an internal face (resp. the outer face) of $G$, where $|f|$ is the length of $f$;
    \item each node of $\cal N$ corresponding to a vertex of $G$ has an outgoing arc directed toward the nodes corresponding to its incident faces; and
    \item each arc of $\cal N$ has a lower bound of $1$ unit of flow.
\end{itemize}
The existence of a flow in $\cal N$ from the sources to the sinks with value $4n$ corresponds to the existence of a rectilinear drawing of $G$ that respects its plane embedding.
It is easy to modify $\cal N$ so that the existence of a flow with value $4n$ corresponds to the existence of a rectangular drawing of $G$. Namely, it suffices to equip each arc of $\cal N$ with an upper bound of $2$ (resp. $3$) if the node of $\cal N$ the arc is incident to corresponds to an internal face (resp. the outer face) of $G$, and with a lower bound of $1$ (resp. $2$) if the node of $\cal N$ the arc is incident to corresponds to an internal face (resp. the outer face) of $G$. The existence of a flow of value $4n$ in $\cal N$ can be tested by using the max-flow~algorithm~of~Borradaile~et~al.~\cite{DBLP:journals/siamcomp/BorradaileKMNW17}. 

If $G$ is not flat, then $G$ is the subdivision of a triconnected planar graph. Let $\cal E$ be the unique planar embedding  of $G$. For each possible selection of a face of $\cal E$ as the outer face, we consider the resulting plane embedding of $G$ and use the same strategy as above to test for the existence of a rectilinear drawing of $G$ that respects such a plane embedding. Since there are $O(n)$ possible choices of the outer face, this results in an $O(n^2 \log^3 n)$-time algorithm.%
\end{proof}

\section{Conclusions and Open Problems}

We studied the recognition of graphs admitting the beautiful drawings that require rectilinear edges of unit length, planarity, and convexity of the faces. We showed that, if the outer face is required to be drawn as a rectangle, the problem is polynomial-time solvable, while it is \NP-hard if the outer face is an arbitrary polygon, even if the input is biconnected, unless such a polygon is specified in advance. These results hold both in the fixed-embedding and in the variable-embedding settings. A byproduct of our results is a polynomial-time algorithm to recognize graphs admitting a rectangular (non-necessarily unit-length) drawing.

It is worth remarking that if the input is a subdivision of a triconnected planar graph, then our algorithms pay an extra time to handle the outer face. Specifically, for unit-length rectangular drawings, an extra quadratic time is used to guess a rectangular drawing of the unique candidate outer face, while, for general rectangular drawings, an extra linear time is used to determine the actual candidate outer face. Hence, it is appealing to study efficient algorithms for this specific case.
Observe that the \NP-hardness results on trees in~\cite{DBLP:journals/ipl/BhattC87,g-ulebt-89} heavily rely on the variable embedding setting.  

\bibliographystyle{splncs04}
\bibliography{biblio-journal}

\appendix

\end{document}